\documentclass{JHEP3}

\usepackage{graphicx,amsmath,amssymb}
\usepackage{epsfig,multicol}
\usepackage{epsf}
\usepackage{epstopdf}

\input{epsf.sty}

\def\nn{\nonumber}

\def\beq{\begin{equation}}
\def\eeq{\end{equation}}
\newcommand{\bea}{\begin{eqnarray}}
\newcommand{\eea}{\end{eqnarray}}

\DeclareMathOperator{\tr}{Tr}

\title{Dual Identities inside the Gluon and the Graviton Scattering Amplitudes}
\author{S.-H. Henry Tye\footnote{sht5@cornell.edu} $\ $  and Yang Zhang\footnote{yz98@cornell.edu}
\\{\em Laboratory for Elementary Particle Physics,
 Cornell University, Ithaca, NY 14853, USA}}
 
\date{\today}

\abstract{
Recently, Bern, Carrasco and Johansson conjectured dual identities inside the gluon tree scattering amplitudes. In this paper, we use the properties of the heterotic string and open string tree scattering amplitudes to refine and derive these dual identities.
These identities can be carried over to loop amplitudes using the unitarity method. Furthermore, given the $M$-gluon (as well as gluon-gluino) tree amplitudes, $M$-graviton (as well as graviton-gravitino) tree scattering amplitudes can be written down immediately, avoiding the derivation of Feynman rules and the evaluation of Feynman diagrams for graviton scattering amplitudes. %The implication of these gluon-graviton duality remains to be explored. 
}


\begin{document}

%\newpage

\section{Introduction}

Gluon and graviton scattering amplitudes have very compact forms which are not obvious at all if one follows the Feynman rules to evaluate the corresponding Feynman diagrams. Using a combination of symmetry, string theory techniques and the spinor helicity formalism, one can simplify the evaluation of these amplitudes considerably. Some years ago, Zhu showed that the terms in the 4-gluon tree amplitude obey an identity \cite{zhu}. Recently, Bern, Carrasco and Johansson conjectured the presence of such identities in higher tree as well as loop amplitudes \cite{BCJ}. If these identities are true, the evaluation of the tree-level $M$-gluon amplitudes can simplify considerably. Furthermore, given the $M$-gluon tree amplitudes, $M$-graviton tree scattering amplitudes can be written down immediately. Loop amplitudes can be obtained from the tree amplitudes using the unitarity method \cite{Unitarity} and these identities can be carried over \cite{BCJ}. In this paper, we use the properties of the heterotic string and open string scattering amplitudes to refine and prove parts of the BCJ conjecture and to extend the identities to include scatterings of massless gluinos and gravitinos.

Consider the $M$-gluon tree scattering amplitude which is a function of the external gluon momenta $k^{\mu}_i$ where $k_i^2=0$ (and $\sum_i k^{\mu}_i=0$), polarizations $\zeta^{\mu}_i$ where $\zeta_i\cdot k_i=0$ and color $a_i$, $i=1, 2, .. M$,
\begin{equation}
\label{decompositionG}
\mathcal A^{\text{YM}}_M(k_i, \zeta_i, a_i) = g^{M-2} \sum_j \frac{c_j(a_i) n_j(k_i, \zeta_i)}{P_j}
\end{equation}
where the sum is over all allowed channels (or terms) with different pole structures. There are $(2M-5)!!$ channels in $\mathcal A^{\text{YM}}_M$. Each denominator is a product of $(M-3)$ pole factors : $P_j= \Pi^{M-3}_{m=1} p_{j,m}(k_i)$, where each pole factor $p_{j,m}(k_i)$ corresponds to the kinematic invariant of an internal gluon propagator. For example, a 2-particle channel pole $p_{j,m}$ takes the form $s_{ln}= -(k_l + k_n)^2$.  
%or $s_{(ln)}= -(k_l +...+ k_n)^2$ for $n>l$. 
The kinematic factor $n_j (k_i, \zeta_i)$ is a function of $k^{\mu}_i$ and $\zeta^{\mu}_i$. Although the choice of the set of $n_j$'s is far from unique, $\mathcal A^{\text{YM}}_M$ itself is independent of the specific choice of $n_j$'s. In this paper, we shall discuss the choices of the $n_j$'s in some detail. The color factor $c_j(a_i)$, a function of the colors $a_i$, is a product of the $(M-2)$ group structure constants $\tilde f^{abc}$ corresponding to the respective pole structure, 
where, for a given Lie algebra, $\tr (T^a T^b)=\delta^{ab},\ [T^a,T^b]=i \sqrt 2 f^{abc} T^c\equiv \tilde f^{abc} T^c$. 
As an illustration, we see that the 5-point diagram in Figure 1 has denominator $s_{13} s_{54}$ and 
$c_{(54)2(13)} =\tilde f^{a_5 a_4 b} \tilde f^{b a_2 d} \tilde f^{da_1 a_3}$. Note that $\mathcal A^{\text{YM}}_M$ is unchanged if we flip the signs of both $c_j$ and $n_j$ ($c_j \to - c_j$ and $n_j \to -n_j$) in any term in Eq.(\ref{decompositionG}). 
At times, we shall set the coupling $g=1$. %(or absorb them into the kinematic factors).

\begin{figure}[h!]
\centering
\includegraphics[scale=0.7]{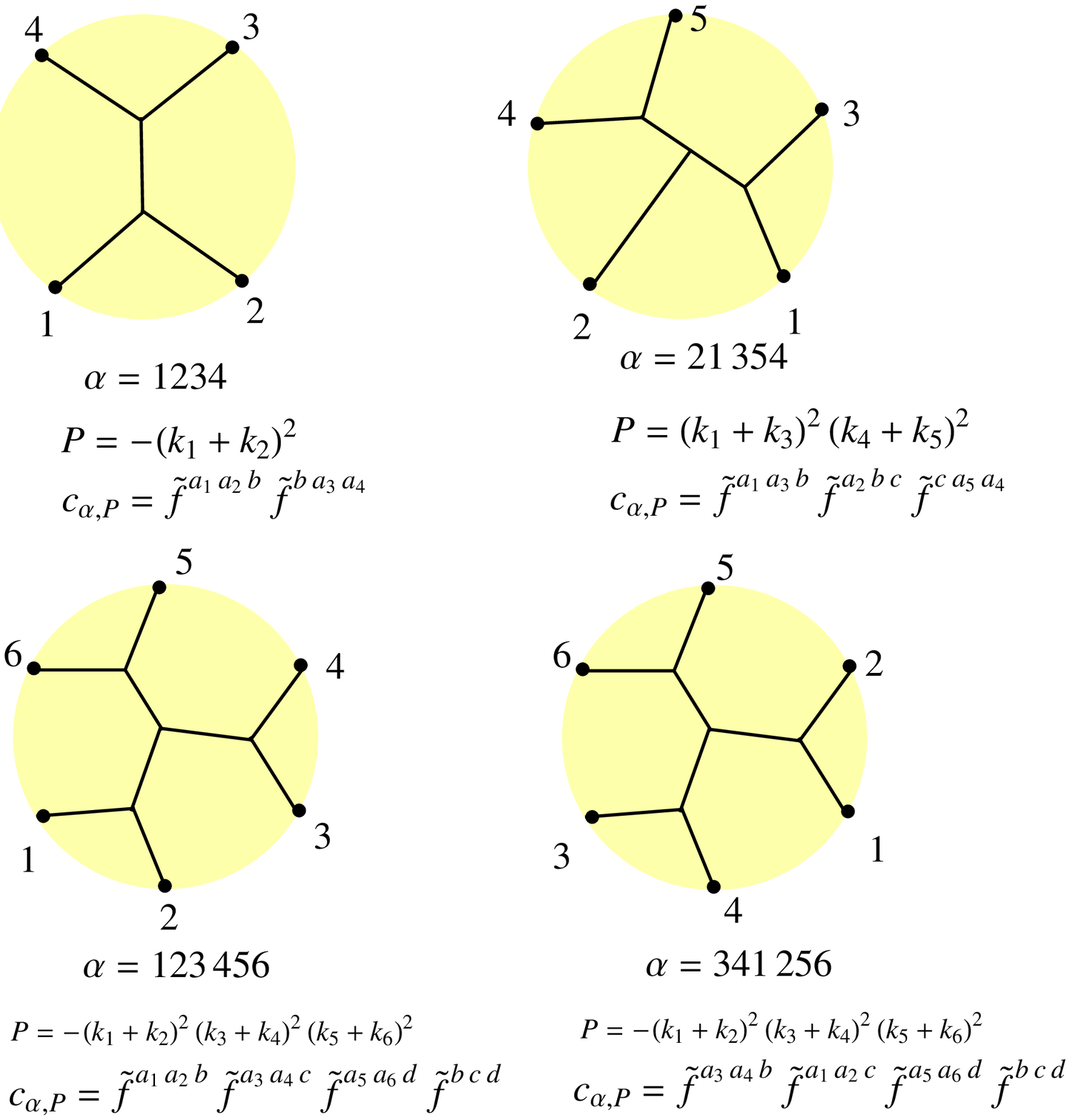}
\caption{Several examples of the poles $P=\Pi_j s_{ln}$ and the color factors. The structure constants are labeled in the counter-clockwise direction. The field theory tree amplitudes $A^{tree}$ are related to the zero-slope limit of the open string amplitudes $A^{open}$, which are given by the disc diagrams in open string theory. The (yellow) disc for each graph is shown to emphasize this feature.} 
\label{color_example}
\end{figure}

The dual identities are best illustrated by the 4-gluon tree level scattering amplitude,
\begin{equation}
\mathcal A^{\text{YM}}_4(k_1,\zeta_1, a_1,...k_4,\zeta_4, a_4)=\frac{c_s n_s}{s}+\frac{c_u n_u}{u}+\frac{c_t n_t}{t},
\label{decomposition}
\end{equation}
where $s, t, u$ are Mandelstam variables, $s= s_{12}= -(k_1+k_2)^2, t=s_{14} = -(k_1 + k_4)^2, u=s_{13} = -(k_1+k_3)^2$ and  $s+t+u=0$. 
Here the color factors
\begin{eqnarray}
c_s &=&\tilde f^{a_1 a_2 b}\tilde f^{b a_3 a_4} \nonumber \\
c_t &=&\tilde f^{a_2 a_3 b}\tilde f^{b a_1 a_4} \nonumber \\
c_u &=& \tilde f^{a_3 a_1 b}\tilde f^{b a_2 a_4}
\label{c_ff}
\end{eqnarray}
depend on the color indices. 
By the Jacobi identity,
\begin{equation}
c_s+c_u+c_t =0.
\label{color_identity}
\end{equation}
It was shown \cite{zhu} that the $n_j (k_i, \zeta_i)$'s satisfy the dual identity,
\begin{equation}
n_s+n_u+n_t=0.
\label{kinematic_identity}
\end{equation}
%This can be checked by looking at the explicit forms $n_j$ given in Appendix {\ref{YMA}. 

Note that $n_{s}(k_i, \zeta_i)$ is determined up to a term proportional to $s$:  
\begin{equation}
n_s \to n'_s=n_s+s \eta (k_i, \zeta_i)
\label{nsgtrans}
\end{equation}
and similarly for $n_u$ and $n_t$, where $\eta$ is an arbitrary function of the kinematic variables. Following Ref.\cite{BCJ}, we shall refer to this as a gauge transformation.
We shall consider only the $\eta$'s that have no pole term, so this ``gauge freedom" changes only the ``contact" part but not the ``residue" or ``non-contact" part of $n_s$. A redistribution of the contact (4-point coupling) term among the 3 terms inside $\mathcal A^{\text{YM}}_4$ (\ref{decomposition}) will lead to such a change in the $n_j$'s.  However, the identity (\ref{kinematic_identity}) is gauge-independent,
\begin{equation}
n'_s+n'_u+n'_t= (n_s+s \eta) + (n_u+u \eta) + (n_t + t \eta) = (s+t+u) \eta =  0. \nonumber
\end{equation}
%In this summation, we used the physical polarization condition $k_i \cdot \zeta_i=0$.
Note that $\mathcal A^{\text{YM}}_4$ is gauge-invariant, as it should be. 

Let us now consider the general $\mathcal A^{\text{YM}}_M(k_i, \zeta_i, a_i)$ (\ref{decompositionG}). Here the $n_j$'s are ``gauge"-dependent, even though $\mathcal A^{\text{YM}}_M$ itself is invariant. Unless specified otherwise, the $n_j$'s are chosen to have no poles, that is, they are local. A convenient symmetrized way of expressing them may be found in Ref.\cite{Cvitanovic:1976am}. There are many triplets of $c_j$ in $\mathcal A^{\text{YM}}_M$ (\ref{decompositionG}) that satisfy
\begin{equation}
c_j +c_l+c_k =0.
\label{color_identityG}
\end{equation}
Each color identity (\ref{color_identityG}) is nothing but the Jacobi identity multiplied by an overall factor of a product of structure constants. One can take any 4 (internal and/or external) gluons in a diagram that are connected by a single internal gluon propagator. The 3 $c_j$'s in a color identity (\ref{color_identityG}) simply correspond to the 3 ways (i.e., the ``$s, t, u$" channels) of connecting those 4 gluons.
BCJ conjectured that whenever a set of 3 $c_j$ in $\mathcal A^{\text{YM}}_M$ (\ref{decompositionG}) satisfy the color identity (\ref{color_identityG}),
the corresponding 3 $n_j$'s in the same $\mathcal A^{\text{YM}}_M$ (\ref{decompositionG}) satisfy the kinematic identity
\begin{equation}
n_j +n_l+n_k =0.
\label{kinematic_identityG}
\end{equation}
In general, there are many such dual pairs of identities for the $M$-point amplitude, not all of them are independent.
One can explicitly check this for the 5-gluon tree amplitude with its 9 independent kinematic identities \cite{BCJ, Mafra}, which we shall also discuss in some detail.  
%Using the unitarity method, 
BCJ also conjectured that these identities can be carried over to loop amplitudes using the unitarity method \cite{Unitarity}.

There are 2 key properties in the BCJ relation : \\
(1) There is a set of kinematic identities (\ref{kinematic_identityG}) in $\mathcal A^{\text{YM}}_M$ (\ref{decompositionG}) for an appropriate set of $n_j$'s; \\
(2) There is a duality between a color identity (\ref{color_identityG}) and the corresponding kinematic identity (\ref{kinematic_identityG}). \\
In this paper, we shall use the properties of the heterotic string model \cite{Heterotic} to prove the duality property between a color identity (\ref{color_identityG}) and the corresponding kinematic identity for the 
%``residue" or ``non-contact" parts of 
$n_j$'s,
 \begin{equation}
 \left( n_j +n_l+n_k \right) {\big |}_{\text{residue}} =0.
\label{k_idGnc}
\end{equation}
where the ``residue" refers to the residue of the product of the $(M-4)$ poles that are common among the $n_j$, $n_l$ and $n_k$ channels. 

To illustrate the difference between the $M=4$ case and the $M>4$ cases, let us look at a $M=5$ open string amplitude identity which yields the following gauge-independent identity,
%We write down the explicit form for the ``coupled" dual Jacobi identity for arbitrary $M$, which is gauge invariant, like 
\begin{eqnarray}
\frac{n_{(13)(42)5}-n_{2(13)(45)}+n_{(13)4(52)}}{s_{13}}+\frac{n_{1(32)(45)}-n_{(21)3(45)}+n_{(13)(45)2}}{s_{45}}\nonumber\\
\frac{n_{(51)(32)4}-n_{2(51)(34)}+n_{(51)3(42)}}{s_{15}}+\frac{n_{(34)2(51)}-n_{(21)(34)5}+n_{1(34)(52)}}{s_{34}}=0
\label{5im-sidex}
\end{eqnarray}
where $n_{(13)(42)5}$ is the numerator factor of the double pole $s_{13} s_{24}$ in $\mathcal A^{\text{YM}}_5$ (\ref{decompositionG}). The details of the $M=5$ case will be explained in Section 5. Here it suffices to note that the 3 $n_j$'s in any one of the 4 triplets has a common pole which appears in the respective denominator in the constraint (\ref{5im-sidex}).
Actually, the corresponding color factors obey identical relations to the $n_j$'s in Eq.(\ref{5im-sidex}). Since the $c_j$'s have discrete values, $ \left( c_j + c_l + c_k \right){\big |}_{\text{residue}} =0$ implies the color identities (\ref{color_identityG}). This is not the case for the $n_j(k_i, \zeta_i)$'s because the momenta $k_i$ are continuous and because of the gauge freedom.  %We shall see that these regular parts obey identities derived from  properties of open string amplitudes.
 Eq.(\ref{5im-sidex}) only implies that the residue of each pole term must vanish. 
For example, the residue of $(n_{(13)(42)5}-n_{2(13)(45)}+n_{(13)4(52)})$ must vanish, but its regular component that is proportional to $s_{13}$ need not. On the other hand, the 4 regular pieces in Eq.(\ref{5im-sidex}) must sum to zero. This property generalizes to arbitrary $M$. There are $(M-3)!(M-3)$ independent open string identities each of which involves $2^{M-3} (M-3) {(2M-7)!!}/{(M-2)!}$ triplets, where each triplet of $n_j$'s is the numerator of a product of $(M-4)$ poles that are common to the $n_j$ channels in that triplet. This yields the set of kinematic identities (\ref{k_idGnc}), in one-to-one correspondence to the color identities (\ref{color_identityG}). 

As conjectured in Ref.\cite{BCJ}, the kinematic identities (\ref{kinematic_identityG}) for $M > 4$ hold only in specific gauge choices. In proving this for $M=5$, we reveal the underlying gauge choice issue. For larger $M$, we support (but do not prove) this part of the BCJ conjecture, that there always exist gauge choices such that (\ref{kinematic_identityG}) holds for the complete set of the kinematic identities. If true, the space of such gauge choices will have dimension $(M-3)!(M-3)$. On the other hand, the kinematic identity (\ref{k_idGnc}) refers to the ``gauge"-invariant part of the $n_j$'s and so may be more relevant. Since the string identities are among gauge-independent partial amplitudes, one should treat them as the defining identities.

Heterotic string model contains gauge fields and their interactions agree with that of the Yang-Mills theory in the zero slope limit (as can be shown in the background field analysis). So their scattering yields tree amplitudes $\mathcal  A^{\text{het}}_{\text{M-gluon}}$ that obey, in the zero Regge slope limit $\alpha ' \to 0$,
\begin{equation}
%\lim_{\alpha ' \to 0} 
\mathcal  A^{\text{het}}_{\text{M-gluon}}(\alpha'=0) = \mathcal A_M^{\text{YM}}
\label{hetYM}
\end{equation}
The amplitudes $\mathcal  A^{\text{het}}_{\text{M-gluon}}$ are functions of open string amplitudes via the KLT relation \cite{KLT}; these open string amplitudes obey identities that yield both the color identities and the kinematic identities on equal footings. 
As we shall see, the duality between the $c_j$ and the $n_j$ also corresponds precisely to the duality between $M$-gluon and $M$-graviton scattering amplitudes, proving yet another BCJ conjecture. The implications of this gauge-gravity duality remain to be further explored. 
In short, we see that there are 2 versions of duality, i.e, a double duality.

Let us briefly review Type I open string theory and explain first why heterotic string theory helps. We then summarize its key properties relevant for showing the duality property. 
If we treat the gluon field as a matrix, $A_{\mu} = A_{\mu}^a T^a$ in perturbation expansion, we obtain
the $M$-gluon tree scattering amplitude as a sum of gauge invariant sub-amplitudes \cite{TreeColor},
\begin{equation}
\mathcal A_M^{\text{YM}} = g^{M-2} \sum_{\sigma \in {S^M/Z_M}} \tr(T^{a_{\sigma_1}} T^{a_{\sigma_2}}T^{a_{\sigma_3}}. . .T^{a_{\sigma_M}} ) A^{tree}(\sigma_1\sigma_2\sigma_3....\sigma_M) 
\label{Asub_amplitude1}
\end{equation}
where $S^M$ is the set of all permutations of $M$ lines, and $Z_M$ is the subset of cyclic permutations that preserve the color trace.
The sum over the set $S^M/Z_M$ is over all distinct cyclic orderings in the trace. 
%where $\sigma$ is a permutation of $(1,2,...M-1)$. 
The color-ordered sub-amplitudes $A^{tree}$ are the 
partial amplitudes that receive contributions from diagrams with a particular cyclic ordering of the $M$ external gluons, so the poles occur only in a limited set of momentum channels made out of sums of cyclically adjacent momenta. They also satisfy the cyclic and the reflection properties,
\begin{equation}
A^{tree}(1,2,3,..., M)= A^{tree}(2,3,...,M,1), \quad \quad A^{tree}(1,2,..., M)= (-1)^M A^{tree}(M,...,2,1)
\label{reflection}
\end{equation}
so there are $(M-1)!/2$ different $A^{tree}$s in $\mathcal A_M$.
Each $A^{tree}$ is gauge-invariant and has $2^{M-2} (2M-5)!!/(M-1)!$ channels, i.e., terms of the form $n_i/P_i$ given in Eq.(\ref{decompositionG}).
It is straightforward to show that $\mathcal A_M$ (\ref{Asub_amplitude1}) is equal to $\mathcal A^{\text{YM}}_M$ (\ref{decompositionG}) %It is easy to show the (\ref{Asub_amplitude}) is equivalent to (\ref{decompositionG}) 
by decomposing each partial amplitude into the channels and calculate the commutators of the matrices. For example, in the 4-gluon case, the terms in (\ref{Asub_amplitude1}) which are related to the $n_s/s$ is, 
\begin{eqnarray}
\tr(T^{a_1} T^{a_2} T^{a_3} T^{a_4} )- \tr(T^{a_2} T^{a_1} T^{a_3} T^{a_4} )-\tr(T^{a_1} T^{a_2} T^{a_4} T^{a_3} )+\tr(T^{a_2} T^{a_1} T^{a_4} T^{a_3}) =c_s
.\end{eqnarray}

Next, we consider Type I open string $M$-gluon tree amplitudes,
\begin{equation}
\mathcal A^{open}_M = g^{M-2} \sum_{\sigma \in {S^{M}/Z_M}} Tr(T^{a_{\sigma_1}} T^{a_{\sigma_2}}T^{a_{\sigma_3}}. . .T^{a_{\sigma_M}} ) A^{open}(\sigma_1 \sigma_2 \sigma_3....\sigma_M) 
\label{openstring}
\end{equation}
where the color properties are contained in the Chan-Paton factor (the trace) while $A^{open}$ is a function of the kinematic variables only. Again, the cyclic and the reflection properties reduce the number of $A^{open}$s from $M!$ to $(M-1)!/2$. Now, relations among the $A^{open}$s follow from the analyticity properties of the open string amplitudes, so, among the $(M-1)!/2$ $A^{open}$s in $\mathcal A^{open}_M$, there are only $(M-3)!$ number of independent ones \cite{KLT}. For a convenient set of the $(M-3)!$ basis amplitudes, we may choose $A^{open}(1, \sigma_2 \sigma_3....\sigma_{M-2}, M-1, M)$, where the first and the last 2 gluon positions are fixed, and the permutations involve the remaining $(M-3)$ gluons  sandwiched between the first and the $(M-1)$th gluon. 
In the zero Regge slope limit,
% ($\alpha' \rightarrow 0$), 
\begin{equation}
 \lim_{\alpha ' \to 0} A^{open} (\sigma_1 \sigma_2 \sigma_3....\sigma_M)  \rightarrow A^{tree}(\sigma_1 \sigma_2 \sigma_3....\sigma_M),
 \end{equation} 
so $\mathcal A^{open}_M$ reduces to the $M$-gluon amplitude $\mathcal A_M$. 
So it follows that there are only $(M-3)!$ number of independent $A^{tree}$s \cite{BDV}.

Consider the 4-gluon tree scattering amplitude in the open string case. In the zero slope limit,
\begin{eqnarray}
A^{tree}({1234})&=&+\frac{n_s}{s}-\frac{n_t}{t}\nonumber \\
A^{tree}({2134})&=&-\frac{ n_s}{s}+\frac{n_u}{u}\nonumber  \\
A^{tree}({1324})&=&-\frac{n_u}{u}+\frac{n_t}{t}.
\label{Idefinition_open}
\end{eqnarray}
which are invariant under the transformation (\ref{nsgtrans}).
Their analyticity properties yield the identities among $A^{open}_4$s. In the zero slope limit, they take the forms \cite{BDV},
%\begin{eqnarray}
\begin{equation}
A^{tree}({1234}) + A^{tree}({2134}) + A^{tree}({1324}) = 0 
\label{open4IdKK}
\end{equation}
\begin{equation}
s A^{tree}({2134}) = t A^{tree}({1324})  \iff  n_s+n_u+n_t=0
\label{open4IdBDV}
\end{equation}
%\end{eqnarray}
The first one is obvious; it is the photon decoupling identity \cite{Mangano:1990by}, or the Kleiss-Kuijf relation for $M=4$ \cite{Kleiss:1988ne}. The second identity yields the kinematic identity (\ref{kinematic_identity}). 
Note that the relations among $A^{open}$ leads to relations among the gauge-invariant partial amplitudes $A^{tree}$. 
%It is easy to check that, u
Using only the relation (\ref{open4IdKK}) and taking the zero-slope limit, we can express the $4$-gluon amplitude from open string theory in terms of $A^{tree}$ (\ref{Idefinition_open}), 
\begin{equation}
\mathcal A_4^{\text{YM}} = c_sA^{tree}({1234}) - c_u A^{tree}({1324})
% \left(\frac{c_s n_s}{s}+\frac{c_u n_u}{u}+\frac{c_t n_t}{t},
\label{opentoYM}
\end{equation}
Using the color identity (\ref{color_identity}), we see that this 
reproduces $\mathcal A^{\text{YM}}_4$ (\ref{decomposition}), as expected. 
For general $M$-gluon amplitudes, the open string amplitudes identities that lead to only $(M-3)!$ independent partial amplitudes \cite{KLT, BDV, Stieberger:2009hq} among the $A^{tree}$s should also produce all the kinematic identities (\ref{k_idGnc}) given above. However, in open string theory, the color properties are in the Chan-Paton factors, so the duality between the color identities and the kinematic identities is not transparent at all.

In the heterotic string theory, on the other hand, there are both compactified dimensions and spacetime dimensions. Discrete momenta in the compactified directions correspond to color, so that the color properties are encoded in the string partial amplitudes. Now the string amplitude identities produce the color identities when we take the momenta in the compactified directions and produce the kinematic identities when we take the momenta in the spacetime directions. So the 2 sets of identities are now on equal footing.
The emergence of one assures the emergence of the other.
%However, the book-keeping of these identities are still complicated to sort out. Fortunately, the color identities (\ref{k_idGnc}) are easy to find (they are simply Jacobi identities). 
This duality property allows us to write down the kinematic identity (\ref{k_idGnc}) corresponding to each color identity (\ref{color_identityG}).
As we shall see, in general, the kinematic identities apply only to the residue part, which is gauge-invariant, but not to the ``contact" part. However, we do believe the BCJ conjecture that there always exists a gauge choice such that the kinematic identities (\ref{kinematic_identityG}) are true.

A couple of comments are in order. Since we are not concerned with the finiteness of the string loop amplitudes, we do not have to restrict ourselves to 10 spacetime dimensions for superstrings (the right-movers of the heterotic string) or to 26 for bosonic strings (the left-movers of the heterotic string). We shall consider gauge groups other than those with even self-dual lattices. To simplify the discussion, we shall restrict our discussion to simply-laced Lie groups, in particular $U(N)$. A key fact we shall use is that the $M$-gluon heterotic tree scattering amplitude $\mathcal  A^{\text{het}}_{\text{M-gluon}}$ equals the Yang-Mills  $M$-gluon tree scattering amplitude in the zero slope limit.
Note that the spectrum in the Type I open string model is very different from that in the heterotic string model. However, both reproduce the $M$-gluon amplitude $\mathcal A^{\text{YM}}_M$ in the zero slope limit. So these 2 sets of stringy properties provide different relations for and insights into $\mathcal A^{\text{YM}}_M$. 

To get a flavor of the properties of $\mathcal  A^{\text{het}}(\alpha'=0)$ from the heterotic string perspective, let us consider the 4-point tree amplitude. The heterotic string theory is a closed string model \cite{Heterotic}.
The KLT relation allows us to write the closed string amplitudes in terms of a sum of products of open string amplitudes \cite{KLT}. Since there is only one ($(M-3)!=1$) independent open string partial amplitude %$A^L$ and one independent $A^R$
 for $M=4$, the 4-point amplitude $\mathcal  A^{\text{het}}_{\text{4}}$ is simply a product of a 4-point open string amplitude $A^L$ for the left-movers (the bosonic string) and an appropriate 4-point open string amplitude $A^R$ for the right-movers (the superstring) multiplied by an appropriate sine factor. In the zero-slope limit (that is, keeping the lowest order in the $\alpha'$ expansion), the sine factor reduces to a Mandelstam variable that removes the double poles present in the product, leaving only single pole terms. More explicitly, we have the 3 left-moving partial amplitudes,
\begin{eqnarray}
A^L_{1234}&=&+\frac{n^L_s}{s}-\frac{n^L_t}{t}\nonumber \\
A^L_{2134}&=&-\frac{ n^L_s}{s}+\frac{n^L_u}{u}\nonumber  \\
A^L_{1324}&=&-\frac{n^L_u}{u}+\frac{n^L_t}{t}.
\label{Idefinition_nL}
\end{eqnarray}
and the 3 right-moving partial amplitudes,
\begin{eqnarray}
A^R_{1234}&=&+\frac{n^R_s}{s}-\frac{n^R_t}{t}\nonumber \\
A^R_{2134}&=&-\frac{ n^R_s}{s}+\frac{n^R_u}{u}\nonumber  \\
A^R_{1324}&=&-\frac{n^R_u}{u}+\frac{n^R_t}{t}.
\label{Idefinition_nR}
\end{eqnarray}
For $i=s, t, u$,  $\mathcal  A^{\text{het}}_{\text{4}}(0)$ becomes \\
(1) the 4-gluon scattering amplitude $\mathcal{A}^{YM}_4$ if $n^L_j=c_j$ are the color factors and $n^R_j=n_j (k_i, \zeta_i)$ are the kinematic factors in (\ref{decompositionG}). In this case, the $A^L$s are simply the partial amplitudes in the scattering of 4 colored (in adjoint representation) massless scalar particles with only cubic couplings; \\
(2) the 4-graviton scattering amplitude, if $n^L_j=n_j (k_i, \xi_i)$ and $n^R_j =n_j (k_i, \zeta_i)$ so the graviton polarization $\epsilon_{\mu \nu}$  is the traceless symmetric part of the product $\xi_{\mu} \zeta_{\nu}$.

The open string amplitude identity (\ref{open4IdBDV}) then yields  
\begin{equation}
n^L_s + n^L_t +n^L_u =0, \quad \quad n^R_s + n^R_t +n^R_u =0
\label{InLnRid}
\end{equation}
The KLT relation tells us that there are 6 equivalent ways to write the full 4-point tree scattering amplitude, using $s+t+u=0$,
\begin{eqnarray}
%\mathcal{A}(1234) 
\mathcal  A^{\text{het}}_{\text{4}}(0)
&=& - s A^L_{1234} A^R_{2134} \left(= \frac{n^L_sn^R_s}{s}  - \frac{n^L_t(n^R_s+n^R_u)}{t} - \frac{(n^L_s + n^L_t)n^R_u}{u} \right) \nonumber  \\
&=& - s A^L_{2134} A^R_{1234} \left(= \frac{n^L_sn^R_s}{s} - \frac{(n^L_s+n^L_u)n^R_t}{t} - \frac{n^L_u(n^R_s + n^R_t)}{u}\right) \nonumber  \\
&=& - t A^L_{1234} A^R_{1324} \left(= - \frac{n^L_s(n^R_t+n^R_u)}{s}  + \frac{n^L_t n^R_t}{t} - \frac{(n^L_s + n^L_t)n^R_u}{u}\right) \nonumber  \\
&=& - t A^L_{1324} A^R_{1234} \left(= -\frac{(n^L_t + n^L_u)n^R_s}{s} + \frac{n^L_t n^R_t}{t}  - \frac{n^L_u(n^R_s+n^R_t)}{u} \right) \nonumber  \\
&=& - u A^L_{2134} A^R_{1324} \left(=   - \frac{n^L_s(n^R_t+n^R_u)}{s} - \frac{(n^L_s + n^L_u)n^R_t}{t}+ \frac{n^L_un^R_u}{u} \right) \nonumber  \\
&=& - u A^L_{1324} A^R_{2134} \left(= - \frac{(n^L_t + n^L_u)n^R_s}{s} - \frac{n^L_t(n^R_s+n^R_u)}{t} + \frac{n^L_un^R_u}{u}\right) 
\nonumber  \\
&=& \frac{n^L_sn^R_s}{s} + \frac{n^L_un^R_u}{u} + \frac{n^L_t n^R_t}{t}
\label{KLT_4pts}
%4KLT6}
\end{eqnarray}
where the identities (\ref{InLnRid}) are used.
This is the 4-gluon amplitude $\mathcal{A}^{YM}_4$ (\ref{decompositionG}) for $n^L_j=c_j$ and $n^R_j=n_j (k_i, \zeta_i)$, $j=s, t, u$. Note that the way $\mathcal{A}^{YM}_4$ (\ref{decompositionG}) is reproduced in the heterotic string approach (\ref{KLT_4pts}) is very different from that in the open string approach (\ref{opentoYM}). 
Here, the specific functions $n_j (k_i, \zeta_i)$ can be extracted from the string theory amplitudes. 
Alternatively, we can start without knowing the identities (\ref{InLnRid}). Demanding that the 6 ways to express $\mathcal  A^{\text{het}}_{\text{4}}(0)$ (\ref{KLT_4pts}) be equal now yields both the identities (\ref{InLnRid}) and the diagonal form (\ref{decomposition}). 
The parallel (or dual) property between the left and the right movers is also clear.  

More generally, open string amplitude identities provide relations among the $(M-1)!/2$ $A_M^L$ so there are only $(M-3)!$ independent $A_M^L$s (similarly for $A^R$s). It is the freedom in choosing the set of independent partial amplitudes that allow us to express  the heterotic M-point scattering amplitudes $\mathcal  A^{\text{het}}_{\text{M}}$ in different but equivalent ways. 
In general, we may obtain the identities (\ref{color_identityG}) and (\ref{kinematic_identityG}) if we use the relation (\ref{hetYM}) and compare (the many equivalent ways of expressing) $\mathcal  A^{\text{het}}_{\text{M-gluon}}(0)$ to $\mathcal A_M^{\text{YM}}$ (\ref{decompositionG}) directly.
Notice that the identities are separate for left-movers and right-movers, that is, the left identity follows from the (left-moving) open string amplitude identities and the right identity follows from the (right-moving) open string amplitude identities.

 It is important to note that the open string amplitude identities do not depend on the explicit forms of $n_j^L$ and $n_j^R$. Choosing spacetime momenta instead of internal discrete momenta, the left (bosonic) amplitudes describe the scattering of massless vector particles, so the same set of (left-moving) open string identities yields the corresponding kinematic identity (\ref{kinematic_identityG}).
% \begin{equation}
%n_{\alpha} + n_{\beta} + n_{\gamma} = 0
%\end{equation}
Since the right (superstring) amplitudes also describe the scattering of massless vector particles, we have, in the zero slope limit, the same functional forms for $n^R_j$ and $n^L_j$, 
\begin{equation}
n^R_j = n^L_j =  n_j
\end{equation}
Now, the open string amplitude identities do not care about the explicit form of the kinematic factor $n_j^L$ or $n_j^R$,
so we can generalize the corresponding set of relations (\ref{k_idGnc}) or (\ref{color_identityG},\ref{kinematic_identityG}) to
\begin{equation}
n^R_j + n^R_l + n^R_k = 0, \quad \quad n^L_j+ n^L_l + n^L_k= 0
\label{gkid}
\end{equation}
where at least the residue parts must hold.
In general, each set of these identities are not necessarily independent, so we are free to select a subset of them as an independent set.
 
Although the open string identities are among the gauge-invariant partial amplitudes, these kinematic identities (\ref{gkid}) (for $M>4$) are gauge-dependent (that is, they are true only in specific gauges).
% because individual $n_j$ are gauge-dependent. We shall give a prescription how to construct them. 
This suggests that the open string amplitude identities among the gauge-invariant partial amplitudes may be more useful in general. In some applications, the knowledge of the existence of the kinematic identities is already sufficient. Although we are not able to prove this part of the BCJ conjecture, we do believe that there always exists a gauge choice where the complete set of kinematic identities (\ref{gkid}) are exact.
%Extracting $n_j$'s from string theory, this issue does not even arise.  

In summary, the open string amplitude identities hold for general $n^L_j$ and separately for $n^R_j$. Using Eq.(\ref{hetYM}), we see that, 
in the zero slope limit, $\mathcal {A}^{\text{het}}_{\text{M}}$ has the diagonal form for $n^L_i=c_j$ and $n^R_j=n_j$, 
%then in general, 
%Once we have 
%This is a formal way to show that the general full amplitude takes the form
\begin{equation}
\label{generalM}
%\mathcal {A}(12...M) 
\mathcal  A^{\text{het}}_{\text{M}}(0)= \sum_j \frac{n^L_jn^R_j}{P_j}
\end{equation}
So, given the $M$-gluon amplitude $\mathcal A^{\text{YM}}_M$ (\ref{decompositionG}), the $M$-graviton amplitude ${A}^{grav}_M$
can be written down immediately by replacing $c_j$ by $n_j(k_i, \xi_i)$ (more accurately, $c_j \to \alpha' n_j(k_i, \xi_i)$ and keeping the lowest order in $\alpha'$), where $\xi^{\mu}_i$ are a new set of polarizations ($\xi_i\cdot k_i=0$),
\begin{equation}
\label{mgraviton}
\mathcal {A}^{grav}_M(k_i, \epsilon_i) = \sum_j \frac{n_j(k_i, \xi_i)n_j(k_i, \zeta_i)}{P_j}
\end{equation}
where the graviton tensor polarizations $\epsilon^{\mu \nu}_i$ is given by the $\mu \nu$-symmetrized product $\xi^{\mu}_i\zeta^{\nu}_i$.
This form of $\mathcal {A}^{grav}_M$ is also conjectured by BCJ.
We can also incorporate massless fermions $f_i$ into the right movers, so that $n^R_j=n_j (k_i, \psi, \zeta_i)$ describes the fermion-vector particle scatterings $f+g \rightarrow f+ g + . . . $ and its cross channels. With $n^L_j=c_j$, $\mathcal{A}^{\text{het}}_{\text{M}}$ in the zero slope limit now becomes \\
(1) the gluon scattering ($n^R_j=n_j (k_i, \psi, \zeta_i)$) with gluinos; \\
(2) the graviton-gravitino scattering amplitude when $n^L_j=n_j (k_i, \zeta_i)$ and $n^R_j=n_j (k_i, \psi, \zeta_i)$.
Further generalization to identities in tree scattering amplitudes involving both gluon and gravitons as well as fermions and gravitinos is straightforward. 

The rest of the paper is organized as follows. Section 2 discusses properties of the heterotic string amplitudes in the zero slope limit that are relevant for understanding the identities and the duality property. The gauge choice issue and open string amplitude properties are also reviewed. In Section 3, we focus on the 4-gluon amplitude. In Section 4, we discuss general $M$-gluon amplitude and we illustrate the issues with the 5-gluon amplitude in Section 5. Since some of the subtle issues appear only for $M>4$, the reader may prefer to read
parts of the discussion on the $M=5$ case before the general $M$ case in Section 4. 
Section 6 contains some discussions.
% $M$-graviton and other amplitudes. 
%To set up notations, we review Yang-Mill amplitudes in Appendix \ref{YMA}. 
Some notations are summarized in Appendix A and some details on the emergence of the color factor (Lie algebra) in the heterotic string amplitudes is discussed in Appendix B.

\section{Yang-Mills, Heterotic and Open String Scattering Tree Amplitudes}

The heterotic string theory \cite{Heterotic} is a closed string model that contains the bosonic string in the spacetime $R^{1,D-1} \times \Gamma^N$ as the left-moving part, and the superstring in the spacetime  $R^{1,D-1} $ as the right-moving part. Here the internal discrete momenta span the $N$-dimensional torus $\Gamma^N$ to form a lattice $\lambda^N$. Loop finiteness (modular invariance) requires $D=10$ and $N=16$ with an even self-dual lattice. Since we are not concerned with this important stringy property, we can choose other values of $D$, say $D=4$ here, and $\lambda^N$ for $U(N)$ or $SO(2N)$. 

On the massless level, the left-movers contain the vector modes and the color modes,
%\begin{equation}
%(\text{vector}) + (\text{color modes})
%\end{equation}
where the color modes contain either the discrete momenta $K^I$'s, which correspond to the roots of the Lie algebra, or the polarizations $\zeta^I$ in the lattice $\lambda^N$, which correspond to vectors in the Cartan subalgebra. The right-movers contain the vector modes and the spinor modes. 
%\begin{equation}
%(\text{vector}) + (\text{spinor})
%\end{equation}
We shall use the superscript $(v)$, $(c)$, $(s)$ to denote the vector, color and spinor sectors.

The gluons in the heterotic string are the product of a left-moving color mode and a right-moving vector mode, i.e.,
$(\text{color}) \times (\text{vector})$.
We shall use the fact that, in the zero slope limit, the $M$-gluon tree heterotic scattering amplitudes $\mathcal A^{\text{het}}_{\text {M-gluon}}$ equals the $M$-gluon amplitude in Yang-Mills theory :
$\lim_{\alpha ' \to 0} \mathcal  A^{\text{het}}_{\text {M-gluon}} = \mathcal A_M^{\text{YM}}$.

Recall that $\mathcal A_M^{\text{YM}}$ is gauge-invariant. Let us take a closer look at this issue.
Consider the terms inside the $M$-gluon amplitude (1.1) that have $(M-4)$ common channels (poles), with $\hat P$ as their product. It is easy to convince oneself that there are 3 and only 3 such terms for each choice of $\hat P$, so
 \begin{equation}
\label{G} 
 P_j= {\hat P}s_j, \quad \quad P_k= {\hat P}s_k,  \quad \quad P_l= {\hat P}s_l
 \end{equation}
where $s_j$, $s_k$ and $s_l$ label the the remaining pole in the $c_j$, the $c_k$ and the $c_l$ term respectively. As will be shown later, the corresponding color factors satisfy the color identity $c_j+c_k+c_l=0$.
 Now, under the gauge transformation 
\begin{eqnarray}
\label{H}
n_j &\to& n'_j=n_j + \eta s_{j} \nonumber \\
n_k &\to& n'_k=n_k + \eta s_{k} \nonumber \\
n_l &\to& n'_l=n_l + \eta s_{l}
\end{eqnarray}
where $\eta$ is a local function of $k_i$ and $\zeta_i$, we have 
\begin{eqnarray}
\label{J}
\mathcal A^{\text{YM}}_M &=& \frac{c_j n_j}{P_j} + \frac{c_k n_k}{P_k}  + \frac{c_l n_l}{P_l} + {\text{rest}} = \frac{c_j n_j}{{\hat P}s_j} + \frac{c_k n_k}{{\hat P}s_k}  + \frac{c_l n_l}{{\hat P}s_l} + {\text{rest}} \nonumber \\
&\to& \mathcal A^{' \text{YM}}_M = \frac{c_j n'_j}{{\hat P}s_j} + \frac{c_k n'_k}{{\hat P}s_k}  + \frac{c_l n'_l}{{\hat P}s_l} + {\text{rest}} %\nonumber \\
= \mathcal A^{\text{YM}}_M + \frac{\eta}{{\hat P}} \left( c_j +c_k+c_l \right) = \mathcal A^{\text{YM}}_M
\end{eqnarray}
so we see that $\mathcal A^{\text{YM}}_M$ is invariant under this transformation.
A general gauge transformation of interest here can be decomposed into $(M-3)(2M-5)!!/3$ (not all independent) transformations, each involving a triplet of terms inside $\mathcal A^{\text{YM}}_M$ as in the case just discussed. For the same product $\hat P$ of $(M-4)$ poles, either 2 or 0 terms with $\hat P$ in the denominator appear in each partial amplitude $A^{tree}$. For the partial amplitudes with 2 such terms appearing, these 2 terms always appear with opposite signs (in the sign convention where $c_j +c_k+c_l=0$) so that the gauge terms $\propto \eta$ cancel. So $A^{tree}$ is also gauge-invariant, as it should be.

An $M$-point $L$-loop heterotic string amplitude has only one closed string diagram. The 
KLT relation \cite{KLT} shows that the heterotic string tree scattering amplitude can be written as a sum of terms, each of which is a product of a left-moving tree scattering amplitude, a right-moving tree scattering amplitude and a factor involving only momentum invariants. These left and right tree amplitudes can be expressed as open string amplitudes. 

A typical $M$-point open string tree ordered amplitude is an integral with $M$ Koba-Nielsen variables $x_i$. Mobius invariance allows us to fix any 3 of them, say $x_1=0$, $x_{M-1}=1$ and $x_M=\infty$.
So the ordered $A(1...M)$ takes the form (up to an $x$-independent factor in front)
\begin{equation}
A(1...M) =  \int_0^1 \Pi_{i=2}^{M-2} dx_i  \Theta (x_{i+1}-x_i)  \Pi_{M>j>i\ge 1} (x_j - x_i)^{\alpha' k_i \cdot k_j/2 + m_{ji}}
\end{equation}
where $m_{ji}$ are integers.
Extending any one of the variables from $-\infty$ to $+\infty$ and closing the the contour leads to a vanishing integral. For example,
extending the integration of $x_2$ to $(-\infty , +\infty)$ and closing its contour, we have \cite{Plahte}
\begin{equation}
\int^{\infty}_{-\infty} dx_2 \int_0^1 \Pi_{i=3}^{M-2} dx_i  \Theta (x_{i+1}-x_i)  \Pi_{M>j>i\ge 1} (x_j - x_i)^{\alpha' k_i \cdot k_j/2 + m_{ji}}=0
%\int^1_0 dx_3 \int^1_{x_3} dx_4 ... \int^1_{x_{M-3}} dx_{M-2}......... =0
\end{equation}
%\begin{equation}
 %\int_0^1 \Pi_{i=2}^{M-2} dx_i  \Theta (x_{i+1}-x_i)  \Pi_{j>i} (x_j - x_i)^{2\alpha' k_i \cdot k_j + n_{ji}} =0
%\end{equation}
Now we can break this $x_2$ integral into ordered pieces: $-\infty \to 0$, $0 \to x_3$, $x_3 \to x_4$, ... , and $1 \to \infty$.  Up to a phase, each equals a different ordered open string amplitude. This way, we obtain a relation among the set of $A^{open}$'s.
Extending the other $x_i$ from $-\infty$ to $+\infty$ on other ordered amplitudes yields additional identities, not all of them are independent. 
As a result of these identities, there are only $(M-3)!$ number of independent ordered open string partial amplitudes $A^{open}$'s.  For a convenient set of the basis amplitudes, we may choose $A(1, \sigma_2, \sigma_3,....,\sigma_{M-2}, M-1, M)$, where the first and the last 2 particle positions are fixed, and the permutations involve the remaining $(M-3)$ particles  sandwiched between the first and the $(M-1)$th ones \cite{KLT}. 

%These relations reduces the number of independent $A^{open}$'s to $(M-3)!$. 
In the zero slope limit, the phases drop out in the real part of the integral mentioned above so it yields a relation among the $(M-1)$ ordered amplitudes \cite{BDV}, 
\begin{equation}
A (213...(M-1)M) + A (123...(M-1)M) +  A (132...(M-1)M) + ... + A (13...(M-1) 2 M) = 0
\end{equation}
This and similar relations (real-sid, or the real parts of the open string identities) are known as the Kleiss-Kuijf relations \cite{Kleiss:1988ne}. These real-SID can be used to reduce the number of amplitudes to a smaller set with $(M-2)!$ $A^{tree}$s. This allows one to simplify the sum (\ref{Asub_amplitude}) into $(M-2)!$ terms \cite{{DelDuca:1999rs}},
\begin{equation}
\mathcal A_M = g^{M-2} \sum_{\sigma \in S^{M-2}} \tilde f^{a_1 a_{\sigma_2} x_1} \tilde f^{x_1 a_{\sigma_3} x_2}...\tilde f^{x_{M-3} a_{\sigma_{M-1}} a_M}  A^{tree}(\sigma_1\sigma_2\sigma_3....\sigma_M) 
\label{Asub_amplitude}
\end{equation}
where $S^{M-2}$ is the permutation group for $(2,...M-1)$. Using the Jacobi identity repeatedly, one can show that it is equivalent to (\ref{decompositionG}).

The imaginary part of the above integral yields another identity (im-SID) among all of them except $A(123...(M-1)M)$ which is real to start with. This yields \cite{BDV},
\begin{equation}
(k_2 \cdot k_1) A (213...(M-1)M) -  \sum_{i=3}^{M-1} \bigg(\sum_{j=3}^i k_2 \cdot k_j\bigg) A (13...i,2,(i+1)...(M-1)M) = 0 
\label{im-sidM}
\end{equation}
%\begin{equation}
%- (-1)^{n_{21}}(k_2 \cdot k_1) A (213...(M-1)M) +  (-1)^{n_{23}}(k_2 \cdot k_3) %A (132...(M-1)M) + ... = 0 
%+ A (13...(M-1) 2 M) = 0
%\end{equation}
Extending the other $x_i$ from $-\infty$ to $+\infty$ on other ordered amplitudes yields additional identities, not all them independent. These identities are among gauge-invariant $A^{tree}$s and so are gauge invariant themselves.
As a result of these identities, there are now only $(M-3)!$ number of independent $A^{tree}$s. They form a set of basis amplitudes.

The open string identities from the contour integral of analytic expressions hold for both the left- and the right-moving parts. The residues of the left-moving string identities for the discrete momenta (color factor) will yield the color identities (\ref{color_identityG}). The right-moving string identities for the partial amplitudes have exactly the same form as the left-movers. If we decompose the right-moving partial amplitudes into channels, with numerators $n_j$'s, then the right-moving open string identities just give the identities for $n_j$ (even when the $n_i$'s are not gauge invariant). In particular, this leads to a set of kinematic identities (\ref{k_idGnc}). 
This is summarized in Table 1.
%{\ref{strategy_dual_identity}}.
%, so  this decomposition is not necessary for real calculations. We just use $n_i$'s to show the duality of the Jacobi identity.) Our strategy is summarized in the 

\begin{table}[h!]
\centering
\begin{tabular}{c|p{3.5cm}|c|p{3cm}}
\hline
 & Momenta  & String identity & Comment\\
\hline
Left: $c_i(K^I,\zeta^I)$ &  discrete momenta  & $c_i+c_j+c_k=0$ & color id.\\
\hline
Right: $n_i(k,\zeta)$  & spacetime momenta & $n_i+n_j+n_k=0$ & kinematic id.\\  
\hline
\end{tabular}
\label{strategy_dual_identity}
\caption{Identities inside the $M$-gluon tree scattering amplitudes}
%Strategy for proving BCJ conjecture}
\end{table}

The open string amplitude identities do not depend on the details of the numerator factors in the channel decomposition of the partial amplitudes. They can be the color factors $c_j$ or the kinematic factors $n_j$. When applied to the left-movers, the open string amplitude identities yields the color identities when applied to the internal dimensions, $\lambda^h$, and yields the kinematic identities when applied to the spacetime dimensions. This one-to-one identity enable us to use the Jacobi identity to locate the color identity and hence the corresponding kinematic identities.
Heterotic string also contains the graviton sector, which has both the left-moving and right-moving momenta noncompact and in the spacetime $R^{1,D-1}$. The graviton scattering amplitude can also be calculated by the KLT relation for the heterotic string, and the scheme is summarized as following:

\begin{table}[h!]
\centering
\begin{tabular}{c|p{3.5cm}|c|p{3cm}}
\hline
 & Momenta  & String identity & Comment\\
\hline
Left: $n_i(k,\xi)$ &  spacetime momenta & $n_i+ n_j+ n_k=0$ & kinematic id. \\
\hline
Right: $ n_i(k,\zeta)$  & spacetime momenta &  $n_i+n_j+n_k=0$  & kinematic id.\\  
\hline
\end{tabular}
\label{graviton_dual_identity}
\caption{Identities inside the $M$-graviton scattering amplitudes}
\end{table}
Here $ n_i(k, \xi)$ is simply $n_i(k, \zeta)$ with the polarizations $\zeta_i$ replaced by a new set of polarizations $\xi_i$.
Note that there are 2 sets of dual pairs here: \\
(1) the $c_j$'s and the $n_j$'s in Table 1,  which is present within YM amplitudes and \\
%\ref{strategy_dual_identity}
(2) the  $c_j$'s in Table 1\ and the $n_j$'s in the left-moving sector in Table 2.\\ 
So, if we replace the left-moving amplitude with discrete momenta and polarizations inside the lattice $\lambda^N$ for the Lie algebra (say $SU(N)$) by the left-moving amplitude with spacetime momenta and polarizations, we convert the $M$-gluon scattering amplitude into the $M$-graviton scattering amplitude (up to a factor of $\alpha'^{M-3}$), 
%If we use the gauge dependent factors $n_i$, then it means,
\begin{equation}
\mathcal A^{\text{YM}}_{M} =\sum_i \frac{c_i n_i}{P_i} \iff  \mathcal A^{grav}_{M\text{-graviton}}=\sum_i \frac{n_i(k,  \xi) n_i(k,  \zeta)}{P_i}
\end{equation}

\begin{table}[h!]
\centering
\begin{tabular}{|p{4.5cm} | c |c c c c|}
\hline   
  \# external gluons & $M$ & 4& 5 & 6 & 7\\
  
\hline
  \# channels in $\mathcal A^{\text{YM}}_M =$ \# $ n_j$ & $(2M-5)!!$ & 3 & 15& 105& 945\\
\hline
  \# partial amplitudes $A^{tree}$ &$(M-1)!/2$ & 3 & 12 & 60 & 360\\
\hline
  \# channels in each $A^{tree}$ &$2^{M-2} {(2M-5)!!}/{(M-1)!}$ &2 & 5 & 14& 42 \\
\hline
  \# independent im-SID &  $(M-3)!(M-3)$ & 1 & 4 & 18 &  96\\
    \hline
  \#$A^{tree}$ in a  real-SID &  $M-1$ & 3 & 4 & 5 &  6\\
  \hline
  \#$A^{tree}$ in an  im-SID &  $M-2$ & 2 & 3 & 4 &  5\\
  \hline
  \# triplets in each im-SID &  $2^{M-3} (M-3) {(2M-7)!!}/{(M-2)!}$ & 1 & 4 & 15 & 56 \\
\hline
  \# identities among $n_j$ &  $(M-3) {(2M-5)!!}/{3}$ & 1 & 10& 105& 1260\\
  \hline
  \# indep. kin. identities  & $(2M-5)!! - (M-2)!$ & 1 & 9 & 81 & 825 \\
\hline
  %\# color identities in one contour integral & $\binom {M-1} {3}$ & 1 & 4 & 10 & 20\\
%\hline 
\# independent $n_j$ & $(M-2)!$ & 2 & 6 & 24 & 120\\
\hline
\# basis $A^{tree}$s & $(M-3)!$ & 1 & 2 & 6 & 24\\
\hline
\# of terms in the KLT& $(M-3)![\frac{1}{2}(M-3)]! [\frac{1}{2}(M-3)]!, M \text{odd}$ & 1 & 2 & 12 & 96\\
 relation  & $(M-3)![\frac{1}{2}(M-4)]! [\frac{1}{2}(M-2)]!, M \text{even}$ &  &  &  & \\
\hline
\end{tabular}
%\\ \\
\label{Overall}
\caption{Summary of the counting of the kinematic factors $n_j$ or equivalently the color factors $c_j$. Note that the number of identities among the $n_j$'s are not all independent. Here, real-SID refers to the real part of an open string amplitude identities (equivalent to the  Kleiss-Kuijf relations) and im-SID refers to the imaginary part of an open string amplitude identities \cite{BDV}. The number of $A^{tree}$'s refers to the number before the real-SID and the im-SID. Some entries are already given in Ref.\cite{BCJ, KLT, BDV}.}
\end{table}

Now there are $(M-3)!$ independent left-moving partial amplitudes and $(M-3)!$ independent  right-moving partial amplitudes. Since a heterotic string amplitude is a sum over the product of a left- and a right-moving amplitude, we can express it as a sum over $[(M-3)!]^2$ terms of a left-moving basis amplitude times a right-moving basis amplitude. However, a judicious choice of basis amplitudes can reduce the number of terms in the sum, especially when $M$ is large. The resulting smallest number of terms known is given in Table 3, which also gives a summary of the counting of $n_j$'s and other relevant quantities as well. The counting of $c_j$'s is exactly the same as that for the $n_j$'s. Taking the $(M-3)!$ independent $A^R=A^{tree}$ as the set of basis amplitudes, we can interpret the KLT formula for 
$\mathcal  A^{\text{het}}_{\text{M}}(0)=\mathcal A^{\text{YM}}_M$ as expressing $\mathcal A^{\text{YM}}_M$ as the linear combination of the $(M-3)!$ basis amplitudes $A^{tree}$s.

\section{The 4-Gluon Tree Amplitude from the Heterotic String Model}

As an illustration, we consider the $4$-gluon tree scattering amplitudes in heterotic string model. This is a long path in obtaining the color  identity as well as the kinematic identity. However, its generalization to M-point is straightforward once we see the underlying properties.
Here, we shall take the following steps to prove the BCJ conjecture for $\mathcal A^{\text{YM}}_4$:
\begin{itemize}
\item  We show that the color factor $c_j$'s emerge as the residue of the different channels in the left-moving amplitude in the color sector. Using the contour integral for these left-moving  open string amplitudes, we prove the color identity (\ref{color_identity}) for the $c$'s.
\item When applied to the right-moving open string amplitudes for the vector sector,
%and find kinematic factor $n$'s. Use the same kind of contour integral to 
the same contour integral argument yields the kinematic identity (\ref{kinematic_identity}). In this manner, the kinematic identity (\ref{kinematic_identity}) is dual to the color identity (\ref{color_identity}).
\item Finally, the KLT relation is used to construct the complete 4-gluon amplitude and show its decomposition (\ref{decomposition}). 
Here, the duality between $c_j$ and $n_j$ are manifest. 
\end{itemize}
Another way to see the duality property is to replace the $c_j$'s in the left-movers by the $n_j$ when we go from the compactified space to spacetime. This yields the 4-graviton scattering amplitude.

\subsection{Left-moving amplitudes}

The left-moving amplitude can be thought as open-string amplitudes with four vertex operator with either compact momentum $K^I$ or the Cartan sub-Lie algebra vector $\zeta^I$ instead of the polarization $\zeta^\mu$. It is straightforward to write out the amplitudes for different orderings of the vertex operators,
\begin{eqnarray}
\mathbf A_{2134}^{L(c)}&=&i^2 \cdot co(2134)\cdot \bigg(-\frac{\alpha '}{4 }\bigg) \nn \\
&\cdot & \int_{-\infty}^0  d x_2\ (-x_2)^{\frac{\alpha '}{2} k_1\cdot k_2 + 2\alpha ' K_1 \cdot K_2} (1-x_2) ^{\frac{\alpha '}{2} k_2 \cdot k_3 + 2\alpha ' K_2 \cdot K_3} f(x_2)\\
\mathbf A_{1234}^{L(c)}&=&i^2 \cdot co(1234)\cdot  \bigg(-\frac{\alpha '}{4 }\bigg) \nn \\
&\cdot &\int_{0}^1 d x_2\ x_2^{\frac{\alpha '}{2} k_1 \cdot k_2 + 2\alpha ' K_1 \cdot K_2} (1-x_2) ^{\frac{\alpha '}{2} k_2 \cdot k_3 + 2\alpha ' K_2 \cdot K_3} f(x_2)\\
\mathbf A_{1324}^{L(c)}&=&i^2 \cdot co(1324)\cdot  \bigg(-\frac{\alpha '}{4 }\bigg) \nn \\
&\cdot& \int_1^{\infty} d x_2\ x_2^{\frac{\alpha '}{2} k_1\cdot k_2 + 2\alpha ' K_1 \cdot K_2} (x_2-1) ^{\frac{\alpha '}{2} k_2 \cdot k_3 + 2\alpha ' K_2 \cdot K_3} f(x_2)
\label{3amp3}
\end{eqnarray}
%Here the symbol $\mathcal A$ means string tree amplitude while we will use $A$ as its low energy limit, i.e., Yang-Mills tree amplitude. 
The factor $i^2$ comes from the vertex normalization, while for general $M$-point scattering amplitude, it would be $i^{M-2}$. The normalization factor $(-\frac{\alpha '}{4})$ is included to obtain the correct normalization for the color factors, which are dimensionless. cancel the undeserved overall factors of the $c$'s.
% and finally in the KLT product, we have to compensate this factor. 
The coefficients $co(2134)$ etc. are cocycles for the root lattice,
\begin{eqnarray}
co(2134)=(-1)^{K_2\star K_2+K_3 \star K_1+K_4 \star K_3 +K_4 \star K_1}\\
co(1234)=(-1)^{K_1\star K_1+K_3 \star K_2+K_4 \star K_3 +K_4 \star K_2}\\
co(1324)=(-1)^{K_1\star K_1+K_2 \star K_3+K_4 \star K_3 +K_4 \star K_2}
\end{eqnarray}
which is reviewed in Appendix A.
The function $f(x_2)$ contains $\zeta^I$, the ``polarization'' in the Cartan subalgebra in the color lattice,
\begin{equation}
f(x_2)=\exp\bigg ( \sum_{i>j}\frac{\zeta_i^I \zeta_j^I}{(x_i-x_j)^2}-\sum_{i\not= j} \frac{\zeta_i \cdot K_j}{(x_i-x_j)}  \bigg) \bigg|_{\mbox{multiple-linear}},
\end{equation}
where only the multi-linear terms in $\zeta^I_i$'s are kept. We have already shifted $\alpha '$ to $\alpha '/4$ in order to use the KLT relation later. However, the discrete momentum $K^I$ just appears on the left-moving amplitude, so the exponent like $2 \alpha ' K_1 \cdot K_2$ is not changed by this shift and we can just set $\alpha '=1/2$ for this product in calculations here.

The 3 amplitudes (3.1)-(\ref{3amp3}) are related since we can consider the contour integral,
\begin{equation}
0=\int_{-\infty+i \epsilon }^{\infty +i \epsilon } d x_2\ x_2^{\frac{\alpha '}{2} k_1\cdot  k_2 + 2\alpha ' K_1 \cdot K_2} (1-x_2) ^{\frac{\alpha '}{2} k_2\cdot  k_3 + 2\alpha ' K_2 \cdot K_3} f(x_2).
\end{equation} 
In terms of the string amplitudes, this reads,
\begin{align}
0=(-1)^{  K_1 \cdot K_2} & e^{i \pi(\frac{\alpha '}{2} k_1 \cdot k_2 )} \cdot  co(2134) \mathbf A_{2134}^{L(c)} + co(1234)  \mathbf A_{1234}^{L(c)} \nonumber \\
& +(-1)^{  K_2 \cdot K_3}e^{-i \pi(\frac{\alpha '}{2} k_2 \cdot k_3 )}   co(1324) \mathbf A_{1324}^{L(c)}
\end{align}
However, it is easy to check that 
\begin{equation}
(-1)^{  K_1 \cdot K_2} co(2134)=co(1234)=(-1)^{  K_2 \cdot K_3} co(1324).
\end{equation}
Therefore we get the string identity
\begin{equation}
  e^{i \pi(\frac{\alpha '}{2} k_1 \cdot k_2 )} \mathbf  A_{2134}^{L(c)} + \mathbf A_{1234}^{L(c)} 
 +e^{-i \pi(\frac{\alpha '}{2} k_2 \cdot k_3 )}    \mathbf A_{1324}^{L(c)}=0.
\label{contour_identity}
\end{equation}
In the low energy limit, we have
$\mathbf A_{1234}^{L(c)}\big|_{\alpha ' \to 0}\equiv A_{1234}^{L(c)}$
 etc. Only the massless poles survive in this limit, so we have
\begin{eqnarray}
A_{2134}^{L(c)}&=&-\frac{\tilde c_s}{s}+\frac{c_u}{u}\nonumber  \\
A_{1234}^{L(c)} &=&\frac{c_s}{s}-\frac{\tilde c_t}{t}\nonumber \\
A_{1324}^{L(c)}&=&-\frac{\tilde c_u}{u}+\frac{c_t}{t}.
\label{definition_c}
\end{eqnarray}
The lowest order of (\ref{contour_identity})'s real part (real-SID) yields \footnote{Here the real (imaginary) part means that we choose the real (imaginary) part of the phases $e^{i \pi (\frac{\alpha '}{2} k_1\cdot k_2)}$. Because the left-moving partial amplitudes $A^{L(c)}$  are either pure real or pure imaginary for fixed $M$, this separation of the phases is valid.}  
\begin{align}
  A_{2134}^{L(c)} +  A_{1234}^{L(c)} 
 +  A_{1324}^{L(c)}=0 
\end{align}
which simplifies to the relations of the $c_i$ coefficients,
\begin{equation}
\tilde c_s=c_s, \quad \tilde c_u=c_u, \quad \tilde c_t=c_t.
\end{equation} 
Furthermore, the lowest order of the imaginary part (im-SID) of (\ref{contour_identity})'s gives,
\begin{equation}
s  A_{2134}^{L(c)}=t  A_{1324}^{L(c)}
\end{equation}
which reduces to the Jacobi identity 
%\begin{equation}
$c_s+c_t+c_u=0 $ (\ref{color_identity}).
%\label{color_identity_string}
%\end{equation}

In Appendix B, we explicitly see that 
%\footnote{Here we may not use the orthonormal basis of the Lie algebra, so we distinguish the upper and lower indices. See the appendix A for the detail.}
$c_s= \tilde f^{a_1 a_2 b} \tilde f^{b a_3 a_4 }$, 
$c_u= \tilde f^{a_3 a_1 b} \tilde f^{b a_2 a_4 }$, and 
$c_t= \tilde f^{a_2 a_3 b} \tilde f^{b a_1 a_4 }$.
%\label{c_ff}
%\end{eqnarray} 
So these $c$'s defined in (\ref{definition_c}) are the same as that in \cite{BCJ}. Hence we see that the left-moving amplitude gives the color factors $c$'s and the Jacobi identity (\ref{color_identity}) they satisfy. We readily admit that this is a complicated way to obtain a very simple result. The payoff is in the parallel derivations of the color identity and the kinematic identity, to which we now turn. 

\subsection{Right-moving amplitudes}

The right-moving superstring amplitudes are obtained in the same way as the left-movers.
% part. Now we have no discrete momentum $K^I$, Cartan sub-Lie algebra $\zeta^I$ or the cocycle. Instead, 
Here we introduce the gluon polarizations $\zeta^\mu_i$ and continuous momenta $k^{\mu}_i$,
\begin{eqnarray}
\mathbf { A}_{2134}^{R(v)}&=& \frac{8 i}{ {\alpha '}  ^2} \bigg(\frac{1}{\sqrt 2}\bigg)^{2}\int_{-\infty}^0  d x_2\ (-x_2)^{\frac{\alpha '}{2} k_1\cdot k_2 } (1-x_2) ^{\frac{\alpha '}{2} k_2 \cdot k_3 } \bar f(x_2)\\
\mathbf  { A}_{1234}^{R(v)}&=& \frac{8 i}{ {\alpha '}  ^2} \bigg(\frac{1}{\sqrt 2}\bigg)^{2} \int_{0}^1 d x_2\ x_2^{\frac{\alpha '}{2} k_1 \cdot k_2  } (1-x_2) ^{\frac{\alpha '}{2} k_2 \cdot k_3 } \bar f(x_2)\\
 \mathbf { A}_{1324}^{R(v)}&=& \frac{8 i}{ {\alpha '}  ^2} \bigg(\frac{1}{\sqrt 2}\bigg)^{2} \int_1^{\infty} d x_2\ x_2^{\frac{\alpha '}{2} k_1\cdot k_2 } (x_2-1) ^{\frac{\alpha '}{2} k_2 \cdot k_3 } \bar f(x_2)
\end{eqnarray}
where the $\bar f(x_2)$ contains the polarizations,
\begin{equation}
\bar f(x_2)=\exp\bigg ( \frac{\alpha '}{2}\sum_{i>j}\frac{\zeta_i \cdot \zeta_j}{(x_i-x_j)^2}- \frac{\alpha '}{2}\sum_{i\not= j} \frac{\zeta_i \cdot k_j}{x_i-x_j}  \bigg) \bigg|_{\mbox{multiple-linear}}.
\end{equation}
and we set $x_1=0$, $x_3=1$ and $x_4=\infty$. The overall factor $8i/{\alpha '}^2 $ comes from the sphere amplitude normalization which is the same for arbitrary $M$-point scattering amplitude. As the left-moving amplitude, we already replaced the $\alpha '$ in open string amplitude, by $\alpha '/4$, to match the close string decomposition. The factor $(1/\sqrt{2})^2$ comes from the commutator convention $[T^a, T^b]=i \sqrt 2 f^{abc} T^c$, 
%like the table.({\ref{Color-ordered_Feynman_rules}), 
and exponent ``$2$" in $(1/\sqrt{2})^2$ really means $M-2$ in the general case.

By the same argument for the left-moving part, we have the identity,
\begin{equation}
  e^{i \pi(\frac{\alpha '}{2} k_1 \cdot k_2 )}   \mathbf A^{R(v)}_{2134} +  \mathbf A^{R(v)}_{1234} 
 +e^{-i \pi(\frac{\alpha '}{2} k_2 \cdot k_3 )}  \mathbf A^{R(v)}_{1324}=0.
\label{contour_identity_right_4}
\end{equation}
The right hand amplitude like $A^{(v)}_{1234}$ etc. is just the open string amplitude $A^{open}_{1234}$ with $\alpha '$ replaced by $\alpha '/4$. However, in the zero slope limit $\alpha '\to 0$, the open string amplitudes reduce to the Yang-Mills color-ordered partial amplitudes, which have no dependence on $\alpha '$. Therefore, in the same limit, 
\begin{equation}
\lim_{\alpha ' \to 0} \mathbf A^{R(v)}_{1234}\equiv A_{1234}^{R(v)}=A^{\text{tree}}({1234})
\end{equation}
etc., because the right moving amplitudes have the same forms as the open string amplitudes whose zero slope limit are the Yang-Mills color-ordered partial amplitudes. 
The lowest order in $\alpha '$ of (\ref{contour_identity_right_4}) will determine the identities of the partial amplitude.
\begin{align}
  A_{2134}^{R(v)} +  A_{1234} ^{R(v)} 
 +  A_{1324}^{R(v)}  =0\\
s  A_{2134}^{R(v)} =t  A_{1324}^{R(v)}  \label{imaginary_right_4}
\end{align}
which was obtained in \cite{BDV}. Note that these identities involve only the gauge invariant partial amplitudes.
We can decompose the  partial amplitudes $A$'s into different channels to obtain 
%show the dual BCJ identities explicitly,
\begin{eqnarray}
 A_{2134}^{R(v)} &=&-\frac{ n_s}{s}+\frac{n_u}{u}\nonumber  \\
 A_{1234}^{R(v)} &=&\frac{n_s}{s}-\frac{ n_t}{t}\nonumber \\
 A_{1324}^{R(v)} &=&-\frac{ n_u}{u}+\frac{n_t}{t}.
\label{definition_n}
\end{eqnarray}
so Eq.(\ref{imaginary_right_4}) yields the kinematic identity $n_s+n_t+n_u=0$ (\ref{kinematic_identity}). 
%\begin{equation}
%n_s+n_t+n_u=0.
%\label{kinetic_identity_string}
%\end{equation}
which is dual to the Jacobi identity $c_s+c_t+c_u=0$. 
Although the partial amplitudes are gauge invariant, the $n_j$ are not. However,  the kinematic identity (\ref{kinematic_identity}) is also gauge-invariant. 
%The decomposition of each $ A$ is clear for most terms, according to their poles. However, the separation of contact terms are a little more difficult. We can use the rules (\ref{Channel_Feynman_rules}) to absorb the contact terms. 
%Unlike the partial amplitude, the $n_i$ are not gauge invariant but
%Again the string amplitude gives the natural definition of $n$'s and also the %same identity as Eq.(\ref{kinematic_identity}).

\subsection{The Yang-Mills amplitude}

Finally, we can use the KLT relation to find the 4-gluon string amplitude and its field theory limit. For 4-gluon amplitude, the KLT relation \cite{KLT} reads,
\begin{equation}
\mathcal A_{\text{4-gluon}}^{\text{het}}=- \pi \bigg(\frac{g}{\pi}
\bigg)^2 \sin\bigg(\pi \frac{\alpha '}{2} k_2 \cdot k_3\bigg)  \cdot \bigg(-\frac{4}{\alpha '}\bigg) \mathbf A_{1234}^{L(c)} \mathbf A_{1324}^{R(v)}.
\end{equation}
The low energy limit can be obtained by keeping the lowest order in $\alpha '$ in each term,
\begin{equation}
%\mathcal A_{4}^{\text{YM}}
\mathcal A_{\text{4-gluon}}^{\text{het}} (0) =-\pi^2 \bigg(\frac{g}{\pi}
\bigg)^2  t A_{L,1234}^{L(c)} A_{1324}^{R(v)}.
\end{equation}
Note that all $\alpha '$ cancel as they should. Using Eq.(\ref{definition_c}) and Eq.(\ref{definition_n}) we obtain
\begin{eqnarray}
\mathcal A_{\text{4-gluon}}^{\text{het}} (0)&=& -\pi^2 \bigg(\frac{g}{\pi}
\bigg)^2 t(\frac{c_s}{s}-\frac{c_t}{t}) (-\frac{n_u}{u}+\frac{n_t}{t})\\
%&=& \pi^2 \kappa^2  \bigg((-u-s) \frac{c_s n_u}{s u}-\frac{c_t n_u }{u}-\frac{c_s n_t}{s} +\frac{c_t n_t}{t} \bigg)\\
%&=& \pi^2 \kappa^2  \bigg((-\frac{c_s n_u}{s }-\frac{c_s n_t}{s})+(-\frac{c_s n_u }{u}-\frac{c_t n_u}{u}) +\frac{c_t n_t}{t} \bigg)
&=& g^2 \bigg(\frac{c_s n_s}{s}+\frac{c_u n_u}{u} + \frac{c_t n_t}{t} \bigg),
\end{eqnarray}
where the identities (\ref{color_identity}), (\ref{kinematic_identity}) and $s+t+u=0$ are used.
%\begin{equation}
%\mathcal A_{\text{4-gluon}}^{\text{YM}} = \pi^2 \kappa^2  \bigg(\frac{c_s n_s}{s}+\frac{c_u n_u}{u} + \frac{c_t n_t}{t} \bigg),
%\label{KLT_4pts}
%\end{equation}
This is the 4-gluon amplitude $\mathcal A_{4}^{\text{YM}}$ (\ref{decomposition}) or (\ref{KLT_4pts}).

\section{M-gluon Tree Scattering Amplitudes}

The above analysis generalizes to the $M$-gluon amplitudes. However, for $M>4$, the subtle issue of gauge dependence emerges, complicating the analysis.
%The key to the duality relation between the color identities and the kinematic identities may be summarized as
%(1) both follow from the open  when applied to the left-movers and the right-movers of the heterotic string. 
In this section we consider the $M$-gluon heterotic string tree amplitude that yields the corresponding Yang-Mills scattering amplitude,
\begin{itemize}
\item The heterotic string left-moving amplitudes for the color sector introduces the $c_i$'s, which are the color factors. By the contour integral argument, $c_i$'s satisfy linear identities, which are shown to be the color (Jacobi) identities (\ref{color_identityG}). The discreteness of the internal momenta makes the generalization to general $M$ straightforward. 
\item The heterotic string right-moving amplitudes for the vector sector introduces the kinematic factors $n_j$'s. The same contour integral arguments yield the identities among the $A^{tree}$s, and can be expressed into identities among the $n_j$'s. Due to the continuous nature of the spacetime momenta, the gauge dependence issue needs a more careful treatment. In particular, we obtain a refined version of the BCJ conjecture \cite{BCJ}, namely the kinematic identities (\ref{k_idGnc}).
\item By the KLT relations, the complete amplitude is a product of left and right moving-amplitudes, which reproduces the $M$-gluon amplitude via the relation (\ref{hetYM}).
%\begin{equation}
%A_\text{M-gluon}^{\text{YM}}=\sum_{i} \frac{c_i n_i}{P_i}
%\end{equation} 
%where $P_i$ are products of the poles of channels with an internal gluon propagator.
\end{itemize} 

%proven for arbitrary M-points by heterotic string method.

\subsection{Left-moving amplitudes}

%\subsubsection{Zero Regge slope limit}

As before, the left-moving amplitude are the scattering amplitude of holomorphic  
vertex operators with spacetime momenta $k^\mu_i$ and color parts (discrete momentum $K^I_i$ or discrete polarization $\zeta^I_i$), but without the spacetime polarization. 

For M-point scattering amplitude, up to cyclic symmetry, there are $(M-1)!$ vertex orderings, say, $\mathbf A_{\sigma_1 \sigma_2...\sigma_{M-1} M}^{L(c)}$, where $\sigma$ is a permutation of the first $M-1$ vertices. Again, its zero slope limit is $A_{\sigma_1 \sigma_2...\sigma_{M-1} M}^{L(c)}$, which will be used for the Yang-Mills scatting amplitude. For the sake of simplicity, we shall use the Greek letters $\alpha$, $\beta$ etc. to represent the vertex order like ``$\sigma_1 \sigma_2...\sigma_{M-1} M$".

Each $A_\alpha^{L(c)}$ contains $2^{M-2} (2M-5)!!/(M-1)!$ channels \cite{BCJ}, 
\begin{equation}
A_\alpha^{L(c)}=\sum_{P} \frac{c_{\alpha,P}}{P},
\label{left-moving_expansion}
\end{equation}
where $P$ is a product of $(M-3)$ poles, going through these different $2^{M-2} (2M-5)!!/(M-1)!$ channels. (Note that the number of channels is $C(M-2)$, where $C(n)$ is simply the $n$th Catalan number.)
 For example, there are two channels for $A_{1234}^{L(c)}$, i.e., $P$ is $s$ or $t$.

For $M$ points, the generalization of Eq.(\ref{c_ff}) is,
\begin{equation}
c_{\alpha,P}=\text{Contraction of the $\tilde f$ s}.
\label{c_color_factor}
\end{equation}
where the r.h.s. is determined by the rules in the channel decomposition subsection, 
\begin{itemize}
\item Draw a color-ordered diagram according to the ordering $\alpha$ and the pole $P$ by using the 3-point vertices only.
\item For each  3-point vertex, read $\tilde f^{abc}$ if $(abc)$ is CCW. For each propagator, read $\delta^{ab}$.
\end{itemize}
Eq.(\ref{c_color_factor}) can be proven for general $M$ by the unitarity relation of the tree amplitude and the induction on $M$. 
%We can break an internal line with the momentum $k$ and then separate a digram into two sub-diagrams with the indices order $\alpha$, $\beta$ and the poles $p_1$, $p_2$ respectively, so the unitarity relation reads
%\begin{equation}
%\frac{c_{\alpha \beta,p}}{p}= \frac{c_{\alpha a,p_1}}{p_1} %\frac{\delta^{ab}}{-k^2} \frac{c_{b \beta,p_2 }}{p_2} 
%\end{equation} 
%where $p=-k^2 \cdot p_1 p_2$, and $a$, $b$ are two color indices.  So,
%\begin{equation}
%c_{\alpha \beta,p}= c_{\alpha a,p_1 }  c_{a \beta, p_2}.
%\end{equation}
%The by the induction on $M$ and simple calculation of the three point diagram, %Eq.(\ref{c_color_factor}) is proven.
As in the $M=4$ case, $c_{\alpha,P}$, which corresponds to the lowest order in $\alpha '$ in string amplitude, does not have spacetime momentum dependence. So there is no contact terms in the left-moving amplitude, and the expansion (\ref{left-moving_expansion}) is well-defined.

It is obvious that two different vertex-operator orderings, $\alpha$ and $\beta$, may contain a common channel and so there exist two factors $c_{\alpha,P}$ and $c_{\beta,P}$ with the same $P$. An example is the two 6-point diagrams in Fig.\ref{color_example} (disc diagrams in open string theory), where $A_{123456}^{L(c)}$ and $A_{341256}^{L(c)}$ contain the common channel which corresponds the same pole $P=-(k_1+k_2)^2 (k_3+k_4)^2 (k_5+k_6)^2$, which are related by $A_{123456,P}^{L(c)}=-A_{341256,P}^{L(c)}$. In cases like this, we have,
\begin{equation}
c_{\alpha,P}=\pm c_{\beta,P},
\label{common_channel}
\end{equation}cases
where a minus sign will appear each time when we flip the two legs of an internal vertex.
The color factors satisfy the color identities which always involve 3 $c$'s. This can be verified either by the explicit formula Eq.(\ref{c_color_factor}) and the Jacobi identity or the contour integral argument similar to the 4-point case, as we shall explain now.
%see in the next subsection.

\subsection{Contour integral method and the color (Jacobi) identities}   

As in the $M=4$ case, we can use the contour integral argument on the left-moving open string tree amplitudes to prove Eq.(\ref{common_channel}) and also all the identities for the color factors. 
%We shall see that the contour integral method gives all the color identities for the color factors.

The M-point open string tree amplitude involves $M-3$ integrals over the Koba-Nielsen variables (vertex operators' positions along the real axis), so there are many ways to use the contour integral argument. Here we just show a particular way which gives all the color (Jacobi) identities.

\begin{figure}[t]
\centering
\includegraphics[scale=0.5]{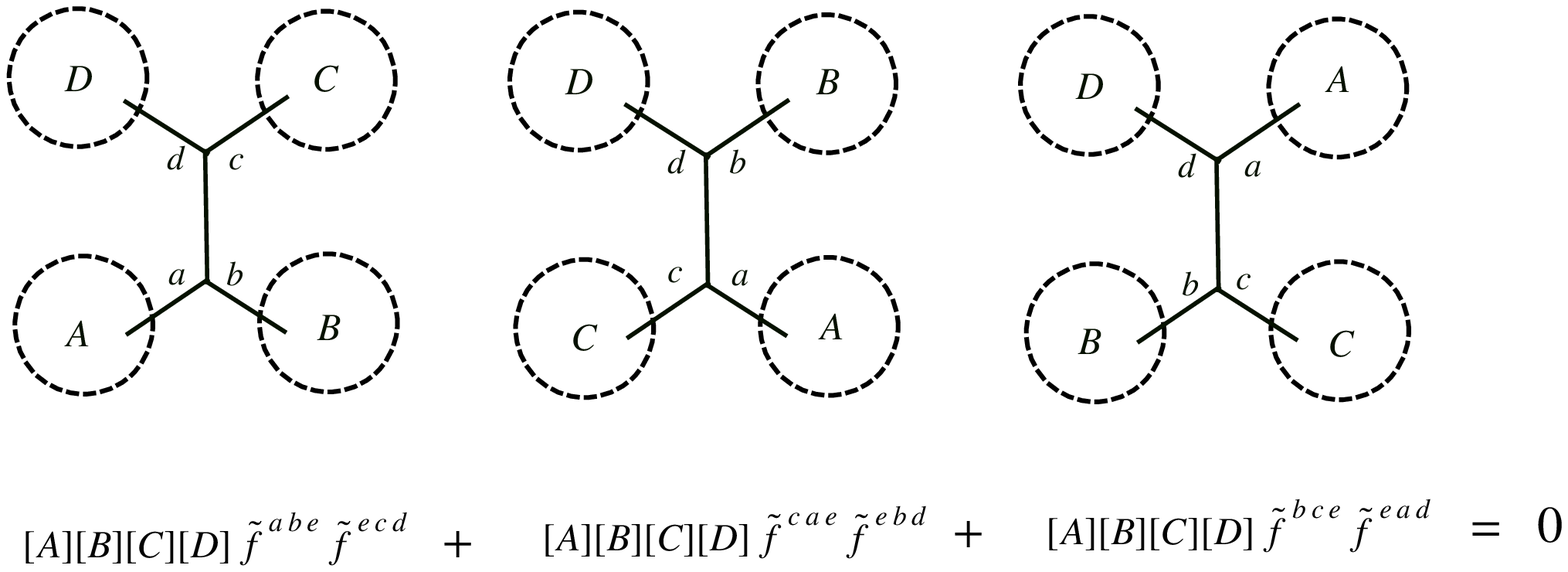}
\caption{General color (Jacobi) identity for the color factors in tree diagrams. The discs A, B, C and D represent the sub-diagrams. }
\label{Jacobi_identity_diagram}
\end{figure}

A general color identity for the color factors, which appears in the tree level amplitude, corresponds to the ``$s$",``$t$",``$u$" channels of four sub-diagrams connected by an internal line. See the diagrams in Fig. \ref{Jacobi_identity_diagram}, where A,B,C,D are four sub-diagrams. [A],[B],[C],[D] are the color factors of the correspondent sub-diagrams, while $a, b, c, d$ are the ``output" color indices of each diagram. The sum of the three diagram's color factors vanishes,
\begin{equation}
[A] [B] [C] [D] (\tilde f^{abe}\tilde f^{ecd}+\tilde f^{bce}\tilde f^{ead}+\tilde f^{cae}\tilde f^{ebd})=0
\label{Jacobi_identity}
\end{equation}
because of the Jacobi identity.
% $\tilde f^{abe}\tilde f^{ecd}+\tilde f^{bce}\tilde f^{ead}+\tilde f^{cae}\tilde f^{ebd}=0$. As the four-point case, Eq.(\ref{Jacobi_identity}) can also be proven by the contour integral argument.

Without loss of generality, we use the following notations for the vertex labels, orderings and poles, \\

\begin{tabular}{ l| l c c}
  Sub-diagram & Vertex ordering & Pole & Color factor\\
\hline
  A & $\alpha=1,...,p$ & P(A) & [A]\\
  B & $\beta=p+1,...,p+q$ & P(B) & [B]\\
  C & $\gamma=p+q+1,...,p+q+r$ & P(C)& [C]\\
  D & $\delta=p+q+r+1,...,p+q+r+s$ & P(D) & [D]
\end{tabular}
\\ \noindent Here $p+q+r+s=M$.

The string amplitude for the vertex ordering $1....M$ is,
\begin{eqnarray}
&& \mathbf A_{1...M}^{L(c)}=i^{M-2}\bigg(-\frac{\alpha '}{4}\bigg)^{M-3} co(1...M)\cdot \int_{x_1< ... < x_{p-1}<0} d x_1 ... d x_{p-1} \int_0^1 d x_{p+1}  \nonumber  \\
& &\int_{x_{p+1}}^1 d x_{p+2}...
\int_{x_{p+q+r-1}}^1 d x_{p+q+r} \int_{1<x_{p+q+r+2}<... x_M} d x_{p+q+r+2}... d x_M
\ s(x) f(x),
\label{string_amplitude_color_M}
\end{eqnarray}
where we fixed $x_{p}=0$, $x_{p+q+r+1}=1$ and $x_M=\infty$. As before,
% the function $s(x)$ is, 
\begin{equation}
s(x)=x_M (x_M-1)\prod_{1\leq i<j\leq M} (x_j-x_i)^{\frac{\alpha '}{2}k_i \cdot k_j +2 \alpha ' K_i \cdot K_j} 
\end{equation}
and 
%the function $f(x)$ is,
\begin{equation}
f(x)=\exp\bigg ( \sum_{1\leq i<j\leq M}\frac{\zeta_i \cdot \zeta_j}{(x_i-x_j)^2}-\sum_{1\leq i\not= j\leq M} \frac{\zeta_i \cdot K_j}{(x_i-x_j)}  \bigg) \bigg|_{\mbox{multiple-linear}},
\end{equation}
again the $\zeta^I_i$'s are the discrete polarizations in the Cartan Lie sub-algebra in the internal compactified space.

Consider the contour-integral of the analytic function $ s(x) f(x)$ in $x_{p+1}$ over the straight line just above the real axis (see Fig. (\ref{contour_integral_diagram})),
\begin{equation}
\int_{-\infty}^{\infty} d x_{p+1} \  \bigg\{i^{M-2} \big(-\frac{\alpha '}{4}\big)^{M-3} co(1...M)\int d x \ s(x) f(x) \bigg\}=0
\label{contour_integral_M}
\end{equation}
where $\int dx$ stands for all the other integrals appearing in (\ref{string_amplitude_color_M}). This equation is similar to the $M=4$ case but with more pieces:

\begin{figure}[t]
\centering
\includegraphics[scale=0.6]{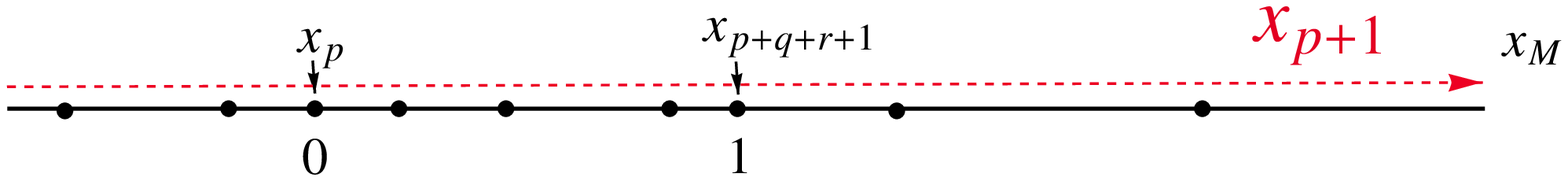}
\caption{The contour integral over $x_{p+1}$.}
\label{contour_integral_diagram}
\end{figure}

\begin{itemize}
\item $0<x_{p+1}<1$. Here, there is only one term which is the original vertex ordering, 
\begin{equation}
\mathbf A_{1...M}^{L(c)}=\mathbf A_{\alpha \beta \gamma \delta}^{L(c)}
\label{ABCD}
\end{equation}
\item $x_{p+1}<0$. Here, the variables $x_{p+2},...,x_{p+q+r}$ are still larger than $x_{p+1}$. Although the orderings inside $\beta \gamma$ and $\alpha$ remain unchanged, the relative ordering between $\beta \gamma$ and $\alpha$ may change. If all the vertex operators in B and C are to the left of A, we have
\begin{equation}
\frac{co(1...M) }{co(\beta \gamma \alpha  \delta)} \mathbf A_{\beta \gamma \alpha  \delta}^{L(c)} \exp \bigg(\frac{i \pi \alpha '}{2} \sum_{i=p+1}^{p+q+r} \sum_{j=1}^{p} k_i \cdot k_j \bigg) (-1)^{2 \alpha ' \sum_{i=p+1}^{p+q+r} \sum_{j=1}^{p} K_i \cdot K_j}.
\end{equation} 
where the ratio between the two co-cycles appears because the two orderings $\alpha \beta \gamma \delta$ and $\beta \gamma \alpha  \delta$ have different cocycles. The exponential term and the power of the $(-1)$ factor come from the changes of the vertex-operator orderings. By repeating the reduction of Eq.(\ref{adjacent}), it is clear that the power of the $(-1)$ factor, which records every adjacent permutation of the vertices, would at the end cancel the ratio of the co-cycles. {\it This cancellation is general for any contour integral.} So this term simplifies to  
\begin{equation}
\mathbf A_{\beta \gamma \alpha  \delta}^{L(c)} \exp \bigg(\frac{i \pi \alpha '}{2} (k_B+k_C) \cdot k_A\bigg)
\label{BCAD}
\end{equation}
where we define
\begin{equation}
\sum_{i=1}^{p} k_i=k_A,\ \sum_{i=p+1}^{p+q} k_i=k_B,\ \sum_{i=p+q+1}^{p+q+r} k_i=k_C.
\end{equation} 
Similarly, when all the vertex operators in $B$ are on the left of that in $A$ and all the vertex operators in $C$ are still on the right of $A$, we have,
\begin{equation}
\mathbf A_{\beta  \alpha  \gamma \delta}^{L(c)} \exp \bigg(\frac{i \pi \alpha '}{2} k_B \cdot k_A\bigg).
\label{BACD}
\end{equation}
For the rest of the terms, the orderings $\alpha$, $\beta$ and $\gamma$ are mixed, for example, and some operators in $\beta$ are inserted into $\alpha$. For these terms, we get,
\begin{equation}
\mathbf A_{\sigma_{1}...\sigma_{p+q+r},p+q+r+1,...,M}^{L(c)} \exp \bigg(\frac{i \pi \alpha '}{2} \sum_{i=p+1}^{p+q+r} \sum_{j=1}^{p} c(i,j;\sigma) k_i \cdot k_j \bigg)
\label{ABC_mixing}
\end{equation}
where $\sigma$ is a permutation of the vertices $\alpha \beta \gamma =1...(p+q+r)$ which keeps intact the relative orderings inside $\alpha$, $\beta$ and $\gamma$, respectively. So $c(i,j;\sigma)=1$ if $\sigma_i<\sigma_j$ and $c(i,j;\sigma)=0$ if $\sigma_i>\sigma_j$.
\item $x_{p+1}>1$. In this case, because the integrals over all the other variables in $\beta$ and $\gamma$ have the original upper bound $1$, we have to reverse their integrals, and obtain
\begin{equation}
(-1)^{q+r-1} \mathbf A_{\alpha, \sigma_{p+1}...\sigma_{M}}^{L(c)}  \exp \bigg(\frac{-i \pi \alpha '}{2} \sum_{i<j}  c(i,j;\sigma)k_i \cdot k_j \bigg)
\label{BCD_mixing}
\end{equation} 
where $\sigma$ is a permutation of the vertices $p+1,...M$. Note that the orderings inside $\beta$ and $\gamma$ are reversed, i.e., for any two indices $i<j$ in $\beta\gamma$, $\sigma_i>\sigma_j$. Since we have fixed $x_M=\infty$, these terms have mixed indices, with the vertices in $\beta$ are inserted into $\delta$.
\end{itemize}  
In summary, the contour integral identity (\ref{contour_integral_M}) takes the form
%\begin{equation}
%(\ref{ABCD})+(\ref{BCAD})+(\ref{BACD})
%+(\ref{ABC_mixing})+(\ref{BCD_mixing})=0
%\label{contour_identity_M}
%\end{equation}
\begin{equation}
\mbox{Eq.(\ref{ABCD})}+\mbox{Eq.(\ref{BCAD})}+\mbox{Eq.(\ref{BACD})}
+\mbox{Eq.(\ref{ABC_mixing})}+\mbox{Eq.(\ref{BCD_mixing})}=0
\label{contour_identity_separated}
\end{equation}
which is the generalization of Eq.(\ref{contour_identity}). Again, as in Eq.(\ref{contour_identity}), the co-cycles do not appear explicitly in the identity. 

%Hence, although there is no co-cycle on the right-moving amplitude, we can still use the contour integral method for the right-moving amplitude to get an analogous identity.

In the zero slope limit, the real part of Eq.(\ref{contour_identity_separated}) just gives Eq.(\ref{common_channel}). For example, Eq.(\ref{contour_identity_separated}) have two terms which contain the pole $P\equiv-p(A)p(B)p(C)p(D)(k_A+k_B)^2$,
\begin{equation}
c_{\alpha \beta \gamma \delta,P}=-c_{\beta \alpha \gamma \delta,P} 
\end{equation} 
where the first terms comes from the amplitude $A_{\alpha \beta \gamma \delta}^{L(c)}$ while the second one comes from $A_{\beta \alpha \gamma \delta}^{L(c)}$ in the zero slope limit. 

%This result chooses the minus sign in Eq.(\ref{common_channel}) because $c_{\alpha \beta \gamma \delta,p}$ and $c_{\beta \alpha \gamma \delta,p}$ are related by a flip of the two sub-diagrams $A$ and $B$.

The imaginary part of Eq.(\ref{contour_identity_separated}) gives the color identities in the zero slope limit. The calculation is similar to the 4-point case, although we need to be more careful since the internal momenta may be off shell, i.e., $k_A^2\not =0$. For the zero slope limit, each one has the form, $k^2 A^{(c)}$. So we are not looking at the poles $-P(A)P(B)P(C)P(D)(k_A+k_B)^2$ of the order $k^{2(M-3)}$ but that of the order $k^{2(M-4)}$, say, $P'=p(A)p(B)p(C)p(D)$. The mixing terms (\ref{ABC_mixing}) and (\ref{BCD_mixing}) cannot give this pole when multiplied by a $k^2$ term. So the relevant terms are just from (\ref{BCAD}) and (\ref{BACD}), $A_{\beta \gamma \alpha \delta}^{L(c)} (k_A\cdot k_B +k_A \cdot k_C)$, $ 
A_{\beta \alpha \gamma \delta}^{L(c)} k_A \cdot k_B$ which gives the six possible terms,
\begin{eqnarray}
c_{\beta \gamma \alpha \delta, -(k_B+k_C)^2 P'} \frac{k_A \cdot k_B}{-(k_B+k_C)^2 P'},\
c_{\beta \gamma \alpha \delta, -(k_B+k_C)^2 P'} \frac{k_A \cdot k_C}{-(k_B+k_C)^2 P'}\\
c_{\beta \gamma \alpha \delta, -(k_A+k_C)^2 P'} \frac{k_A \cdot k_B}{-(k_A+k_C)^2 P'}, \
c_{\beta \gamma \alpha \delta, -(k_A+k_C)^2 P'} \frac{k_A \cdot k_C}{-(k_A+k_C)^2 P'}\\
c_{\beta \alpha \gamma  \delta, -(k_A+k_B)^2 p'} \frac{k_A \cdot k_B}{-(k_A+k_B)^2 P'}, \
c_{\beta \alpha \gamma  \delta, -(k_A+k_C)^2 p'} \frac{k_A \cdot k_B}{-(k_A+k_C)^2 P'}
\end{eqnarray}
where the third term cancels the last term, i.e, $c_{\beta \gamma \alpha \delta, -(k_A+k_C)^2 p'} =-c_{\beta \alpha \gamma  \delta, -(k_A+k_C)^2 p'}$. For the first two terms, we can rewrite the momentum invariants  as,
\begin{equation}
k_A \cdot k_B+k_A \cdot k_C=-\frac{1}{2} k_A^2+\frac{1}{2} k_D^2-\frac{1}{2} (k_B+k_C)^2
\end{equation} 
The last term will cancel the $(k_B+k_C)^2$ in the denominator so we get the expected pole $P'$. The $k_A^2$ term is not involved in this numerator of the pole term because if $k_A$ is on shell, then this term vanishes. If the $k_A$ is off shell, then the $k_A^2$ appearing in $P(A)$ and also $P'$  would be cancelled by this new $k_A^2$ factor and hence the $P'$ pole structure would be changed. In this manner, we further simplify the identity to,
\begin{equation}
-c_{\beta \gamma \alpha \delta, -(k_B+k_C)^2 P'}+ c_{\beta \gamma \alpha \delta, -(k_A+k_C)^2 P'}+c_{\beta \alpha \gamma  \delta, -(k_A+k_B)^2 P'}=0
\label{Jacobi_string}
\end{equation}
which is the color (Jacobi) identity (\ref{Jacobi_identity}) by Eq.(\ref{c_color_factor}). Since the sub-diagrams $A$, $B$, $C$, $D$ are completely general, we obtain all the possible color identities in the tree level by this analysis.
% contour integral method. However, when the identity (\ref{contour_identity_separated}) is complete calculated, we have not only one color identity (\ref{Jacobi_string}) but many identities at once.
%$\binom {M-1} {3}$ identities at once. 
There are $(2M-5)!!$ channels and each channel contains $(M-3)$ internal lines, so there are $(2M-5)!! (M-3)$ choices of (A)(B)(C)(D). Each color identity involves $3$ terms, so there are $(2M-5)!! (M-3)/3$ color identities.

%In real calculation, when we consider the imaginary part of (\ref{contour_identity_M}) and pick up all the $k^{2(M-4)}$ poles, we only just get the Jacobi identity (\ref{Jacobi_string}) but also many other Jacobi identities.
 
We now have the counting given in Table \ref{Overall} ($n_j$'s replaced by $c_j$'s), 
where the number of $c_i$ means the $c_i$ identified by Eq.(\ref{common_channel}) but before the consideration of the color identities. The number of $c_i$ is calculated from choosing one internal line of a tree diagram.
% however, they may not be independent. 
An independent set may be chosen as \cite{DelDuca:1999rs}
\begin{equation}
f^{a_1 a_{\sigma_2} x_1}{f^{x_1 a_{\sigma_3}x_2}}... f^{x_{M-3}a_{\sigma_{M-1}a_M}}
\end{equation}
where $\sigma$'s are the $(M-2)!$ permutations of $(2,3,...M-1)$. 
So there are $(M-2)!$ linearly independent $c_i$'s.

\subsection{Right-moving amplitudes}

Here we choose the right-moving vertex-operators from the vector sector, so for example, 
the string amplitude for the vertex ordering $1....M$ is,
\begin{eqnarray}
&& \mathbf A_{1...M}^{R(v)}=\frac{8 i}{\alpha '^2}\bigg(\frac{1}{\sqrt{2}}\bigg)^{M-2} \cdot \int_{x_1< ... < x_{p-1}<0} d x_1 ... d x_{p-1} \int_0^1 d x_{p+1}  \nonumber  \\
& &\int_{x_{p+1}}^1 d x_{p+2}...
\int_{x_{p+q+r-1}}^1 d x_{p+q+r} \int_{1<x_{p+q+r+2}<... x_M} d x_{p+q+r+2}... d x_M
\ \bar s(x) \bar f(x),
\label{string_amplitude_vector_M}
\end{eqnarray}
where again we fixed $x_{p}=0$, $x_{p+q+r+1}=1$ and $x_M=\infty$. As before the function $\bar s(x)$ is, 
\begin{equation}
\bar s(x)=x_M (x_M-1)\prod_{1\leq i<j\leq M} (x_j-x_i)^{\frac{\alpha '}{2}k_i \cdot k_j} 
\end{equation}
and the function $\bar f(x)$ is,
\begin{equation}
\bar f(x)=\exp\bigg (\frac{\alpha ' }{2}  \sum_{1\leq i<j\leq M}\frac{\zeta_i \cdot \zeta_j}{(x_i-x_j)^2}-\frac{\alpha ' }{2}  \sum_{1\leq i\not= j\leq M} \frac{ \zeta_i \cdot k_j}{x_i-x_j}  \bigg) \bigg|_{\mbox{multiple-linear}},
\end{equation}
where the $\zeta_i$'s are the gluon polarizations. 

%As the four gluon case we calculated, for $M$-gluon scattering case, 

The right-moving amplitudes $\mathbf A^{R(v)}(k_i, \zeta_i)$ in the general $M$ case
%, which contains the momenta and gluon polarizations, will 
satisfy the same open string identities as the $\mathbf A^{L(c)}$'s. More specifically, the contour integral method for the right-moving part will yield similar equations like Eq.(\ref{contour_integral_M}), with all the  $\mathbf A^{L(c)}$'s replaced by $\mathbf A^{R(v)}$. In the zero slope limit, 
\begin{equation}
\mathbf A_{a_{\sigma_1},a_{\sigma_2},...,a_n}^{R(v)} \to  A_{a_{\sigma_1},a_{\sigma_2},...,a_n}^{R(v)} =A^{tree}(a_{\sigma_1},a_{\sigma_2},...,a_n)
\end{equation}
which are the Yang-Mills partial amplitudes.  Therefore, the real (real-SID) and imaginary (im-SID) part of the string identities will generate the set of relations for the partial amplitudes. The key difference between color and kinematic identities become clear for $M>4$. 

Let us take a closer look at the general im-SID formula.
We begin with the partial amplitude $A_{123...M-1,M}$ and consider the contour integral in $x_2$ while the relative order of $(13...M-1,M)$ are fixed. This yields the relation (\ref{Asub_amplitude}), 
\begin{equation}
(k_2 \cdot k_1) A (213...(M-1)M) -  \sum_{i=3}^{M-1} \bigg(\sum_{j=3}^i k_2 \cdot k_j\bigg) A (13...i,2,(i+1)...(M-1)M) = 0 
\label{contour_integral_M_2}
\end{equation}
Note that this contour is the same as (\ref{contour_integral_M}) if we set $\beta=2$. However, unlike the discussion following (\ref{contour_integral_M}), here we cannot simply pick up the residues but have to work out all the terms inside this open string identity and find the relative signs. Besides $\beta$, the remaining 3 vertex orderings $\alpha$, $\gamma$ and $\delta$, which correspond to the sub-diagram $A$, $B$ and $C$, would combine into $\alpha \beta \gamma \delta=12...M-1,M$, up to cyclic permutations. By the cyclic permutation, we can always put the index ``$1$'' inside $\alpha$, so $\alpha=...1,3,...i$.

These identities contain only the gauge independent amplitudes $A^{R(v)}$ so are very convenient for the KLT relation based on the left-moving $A^{L(c)}$ and right-moving $A^{R(v)}$. If we want to show that the dual kinematic identities have the same form as the color identities, we need to decompose the $A^{R(v)}$ into different channels,
\begin{equation}
A^{R(v)}_\alpha=\sum_{P} \frac{n_{\alpha,P}}{P}
\label{ARtree}
\end{equation}
where $P$ goes through all the $2^{M-2} {(2M-5)!!}/{(M-1)!}$ channels within the ordering $\alpha$. However, the choice of explicit expressions for the $n_{\alpha,P}$'s is not unique. (We may start with a particular choice, like the symmetric way according to the color factors \cite{Cvitanovic:1976am}.) In general, each $n_j$ contains both a ``residue" (or ``non-contact') piece and a ``contact" piece, and only the ``residue" piece will obey the kinematic identities. 

%However, as we shall argue, there always exist a choice of the $n_j$'s such that the dual identity hold exactly.
The terms inside (\ref{contour_integral_M_2}) and related to the decomposition $A$, $B$, $C$ and $D$ are,
\begin{eqnarray}
\big\{k_2\cdot (k_3+...+k_i+k_C)-k_2\cdot(k_3+...+k_i)\big\} \frac{n_{\alpha\gamma 2\delta, -P'(k_2+k_C)^2}}{-P'(k_2+k_C)^2} \nonumber \\
\big\{k_2\cdot (k_3+...+k_i+k_C+k_D)-k_2\cdot(k_3+...+k_i)\big\} \frac{n_{2\alpha\gamma\delta, -P'(k_2+k_A)^2}}{-P'(k_2+k_A)^2} \nonumber \\
\big\{k_2\cdot (k_3+...+k_i+k_C+k_D)-k_2\cdot(k_3+...+k_i+k_C)\big\} \frac{n_{2\alpha\gamma\delta, -P'(k_2+k_D)^2}}{-P'(k_2+k_D)^2} 
\end{eqnarray}
where $P'=P(A) P(C) P(D)$. The related terms inside can be simplified into,
\begin{eqnarray}
 \frac{n_{\alpha\gamma 2 \delta, -P'(k_2+k_C)^2}-n_{2\alpha\gamma\delta, -P'(k_2+k_A)^2}+n_{2\alpha\gamma\delta,-P'(k_2+k_D)^2}}{P(A)P(C)P(D)}.
\end{eqnarray}
Repeat this process, the open string identity (\ref{contour_integral_M_2}) is reduced to the `coupled" dual  identity,
\begin{equation}
\sum_{(A,C,D)}  \frac{n_{\alpha\gamma 2 \delta, -P'(k_2+k_C)^2}-n_{2\alpha\gamma\delta, -P'(k_2+k_A)^2}+n_{2\alpha\gamma\delta,-P'(k_2+k_D)^2}}{P(A)P(C)P(D)}=0
\label{coupled_dual_identity}
\end{equation}
where the sum is over all the possible decomposition $(A,C,D)$ such that $\alpha2\gamma\delta=123...M-1,M$ up to cyclic permutation. Each $(A,C,D)$ gives three terms which have the same form of the color (Jacobi) identities. Because the three sub-diagrams $A$, $C$ and $D$, can be interpreted as a $(M-1)$-point channel with one vertex removed, the number of $(A,C,D)$ is 
\begin{equation}
\frac{2^{M-3}(2M-7)!!(M-3)}{(M-2)!},
\end{equation}
so each open string identity contain ${2^{M-3}(2M-7)!!(M-3)}/{(M-2)!}$ triplets, each of which has the form of the color (Jacobi) identities. Eq.(\ref{coupled_dual_identity}) is gauge invariant and holds in arbitrary choice of the $n$'s. When we pick up the residues of (\ref{coupled_dual_identity}), we get,
\begin{equation}
\big\{n_{\alpha\gamma 2 \delta, -P'(k_2+k_C)^2}-n_{2\alpha\gamma\delta, -P'(k_2+k_A)^2}+n_{2\alpha\gamma\delta,-P'(k_2+k_D)^2}\big\} {\big |}_{\text{residue}}=0
\end{equation}  
However, when $M>4$, the dual identities themselves with the non-residue terms do not hold for an arbitrary choice of the $n$'s.
%Note that we have different ways to do the contour integral and the analysis is essentially the same.
 
Let us give an example to illustrate the difference between the $M=4$ case and the $M>4$ cases. Let us consider the $M=5$ case. %Using Eq.(\ref{coupled_dual_identity}), a specific $M=5$ im-SID 
%We write down the explicit form for the ``coupled" dual Jacobi identity for arbitrary $M$, which is gauge invariant, like 
Using Eq.(\ref{coupled_dual_identity}), we can first draw the disc diagram with only $(1345)$ in CCW direction, so there are two 4-point channels. For each channel, we can remove one of the 2 vertices to get the (A,C,D), so there are four choices of (A,C,D): ((13),4,5), (1,3,(45)),((51),3,4),(1,(34),5). Eq.(\ref{coupled_dual_identity}) yields the gauge-independent identity (\ref{5im-sidex}),
\begin{eqnarray}
0=\frac{n_{(13)(42)5}-n_{2(13)(45)}+n_{(13)4(52)}}{s_{13}}+\frac{n_{1(32)(45)}-n_{(21)3(45)}+n_{(13)(45)2}}{s_{45}}\nonumber\\
\frac{n_{(51)(32)4}-n_{2(51)(34)}+n_{(51)3(42)}}{s_{15}}+\frac{n_{(34)2(51)}-n_{(21)(34)5}+n_{1(34)(52)}}{s_{34}}
\end{eqnarray}
where $s_{13}=-(k_1+k_3)^2$ etc and $n_{(13)(42)5}$ is the numerator factor in the $s_{13}s_{24}$ channel in the partial amplitude $A_{13425}$
%where $s_{12}=-(k_1+k_2)^2$ etc., and the subscript in $n_{(14)(25)}=n_{(25)(14)}$ means it is the numerator factor of the double pole $s_{14} s_{25}$
 in $A^{R(v)}_\alpha$ (\ref{ARtree}) or equivalently in $\mathcal A^{\text{YM}}_5$ (\ref{decompositionG}). The details of the $M=5$ case will be explained in Section 5. Here it suffices to note that the 3 $n_j$'s in any of the 4 triplets has a common pole which appears in the respective denominator.
If we replace the kinematic factors by the color factors, then each triplet of color factors must sum to zero. That is how the color identities appear. However, since the spacetime momenta are continuous, this im-SID only implies that the residue of each pole term must vanish. 
%For example, the residue of $(-n_6+n_8+n_9)$ must vanish, but not its regular component. On the other hand, the 4 regular pieces in the im-SID must sum to zero. This generalizes to arbitrary $M$. 
For general $M$, each im-SID involves a set of triplets, where each triplet of $n_j$'s is the numerator of a product of $(M-4)$ poles that are common to the $n_j$'s in that triplet. This yields the set of kinematic identities (\ref{k_idGnc}), in one-to-one correspondence to the color identities (\ref{color_identityG}). 
For $M=5$, we can prove that there always exists a gauge choice so that each triplet sums exactly to zero. We have not been able to extend the proof to general $M$. It is clear that even if such a gauge choice exists, it is hard to find and so the exact identity (\ref{kinematic_identityG}) may not be that useful. On the other hand, the im-SID in terms of the $n_j$'s is gauge-independent and so may be more useful. However, the existence of the exact identity (\ref{kinematic_identityG}) (but without the explicit construction of the $n_j$'s ) can sometimes be useful.

To get a feeling of what to expect as $M$ increases, we give an explicit example for (\ref{coupled_dual_identity}) for the $M=6$ case.
Now we need to draw a 5-point disc diagram with $(13456)$ in CCW direction, which contains $5$ channels. For each channel, we can remove one of the 3 vertices to get a choice (A,C,D), so there are 15 choices, yielding
{
\allowdisplaybreaks
\begin{eqnarray}
0&=&\frac{n_{1((34)2)(56)}-n_{(21)(34)(56)}+n_{1(34)((56)2)}}{s_{34} s_{56}}+\frac{n_{(1(34))(52)6}-n_{(2(1(34)))56}+n_{(1(34))5(62)}}{s_{34} s_{134}}\nonumber \\
&&\frac{n_{((56)1)(32)4}-n_{(2((56)1))34}+n_{((56)1)3(42)}}{s_{56} s_{156}}+\frac{n_{(61)(32)(45)}-n_{(2(61))3(45)}+n_{(61)3((45)2)}}{s_{45} s_{16}}\nonumber\\
&&\frac{n_{((61)3)(42)5}-n_{(2((61)3))45}+n_{(6(13))4(52)}}{s_{16} s_{163}}+
\frac{n_{1((3(45))2)6}-n_{(21)(3(45))6}+n_{1(3(45))(62)}}{s_{45} s_{345}}\nonumber \\
&& \frac{n_{(13)(42)(56)}-n_{(2(13))4(56)}+n_{(13)4((56)2)}}{s_{13} s_{56}}+ \frac{n_{((13)4)(52)6}-n_{(2((13)4))56}+n_{(2((13)4))56}}{s_{13} s_{134}}\nonumber\\
&& \frac{n_{1(32)(4(56))}-n_{(21)3(4(56))}+n_{13((4(56))2)}}{s_{56}s_{456}}+
 \frac{n_{(61)((34)2)5}-n_{(2(61))(34)5}+n_{(61)(34)(52)}}{s_{61} s_{34}}\nonumber\\
&& \frac{n_{(5(61))(32)4}-n_{(2((56)1))34}+n_{(5(61))3(42)}}{s_{16} s_{165}}+ \frac{n_{1(((34)5)2)6}-n_{(21)((34)5)6}+n_{1((34)5)(62)}}{s_{34} s_{345}}
\nonumber\\
&&\frac{n_{(13)((45)2)6}-n_{(2(13))(45)6}+n_{(13)(45)(62)}}{s_{13} s_{45}}+ \frac{n_{(6(13))(42)5}-n_{(2(6(13)))45}+n_{(6(13))4(52)}}{s_{13} s_{136}}
\nonumber\\
&& \frac{n_{1(32)((45)6)}-n_{(21)3((45)6)}+n_{13(((45)6)2)}}{s_{45} s_{456}}
\end{eqnarray}
}
where $s_{156}=-(k_1+k_5+k_6)^2$ etc.. Here, $n_{1((34)2)(56)}$ is the numerator factor in the $s_{34} s_{234} s_{56}$ channel in the partial amplitude $A_{134256}$ in $A^{R(v)}_\alpha$ (\ref{ARtree}). For $M=6$, there are 18 independent im-SID's and each im-SID contains $15$ triplets.

\subsection{Yang-Mills amplitudes}

In this subsection, we determine the Yang-Mills amplitude for $M$ gluons,
% In term of the $c_i$'s and $p_i$'s, the amplitude is
\begin{equation}
A_{\text{M}}^{\text{YM}}= g^{M-2} \sum_{P} \frac{c_P n_P}{P} 
\label{Yang_Mills_Amplitude}
\end{equation}
where $P$ goes through all the $(2M-5)!!$ channels, i.e., all the different pole structures. The product $c_P n_P$ means $c_{\alpha,P} n_{\alpha,P}$, which actually has no dependence of the vertex ordering $\alpha$. A different ordering choice $\beta$ will introduce $\pm$ signs by Eq.(\ref{common_channel}) $c_{\alpha ,P}=\pm c_{\beta ,P}$, $\ n_{\alpha ,P}=\pm n_{\beta ,P}$,
however, they always take the same sign so $c_{\alpha,p} n_{\alpha,p}=c_{\beta,p} n_{\beta,p}$.

We can use the KLT relation \cite{KLT} to derive Eq.(\ref{Yang_Mills_Amplitude}),
\begin{equation}
\mathcal A_{\text{M-gluon}}^{\text{het}}=\bigg(\frac{i}{2}\bigg)^{M-3} \pi \bigg(\frac{g}{\pi}\bigg)^{M-2} \bigg(-\frac{4}{\alpha '}\bigg)^{M-3}\sum_{\alpha,\beta} \mathbf  A_{\alpha}^{L(c)} \mathbf A_{\beta}^{R(v)} F(\alpha,\beta)
\label{KLT_M}
\end{equation}
where $\alpha$, $\beta$ are the $\frac{1}{2}(M-1)!$ vertex orderings since three vertices are fixed. $F(\alpha, \beta)$ is the phase factor,
\begin{equation}
F(\alpha,\beta)=\exp \bigg(i \pi \sum_{i>j} f(\frac{\alpha '}{2} k_i \cdot k_j;\alpha, \beta)\bigg)
\end{equation}
where ($i>j$)
\begin{equation}
f(\frac{\alpha '}{2} k_i \cdot k_j;\alpha, \beta)=
\begin{cases} 
   \frac{\alpha '}{2} k_i \cdot k_j & \text{if }  (\alpha_i-\alpha_j)(\beta_i -\beta_j )<0 \\
   0  & \text{if } (\alpha_i-\alpha_j)(\beta_i -\beta_j )>0
\end{cases}
\end{equation}
so $F(\alpha,\beta)$ is symmetric in $\alpha$ and $\beta$. Contour integral will simplify the expression (\ref{KLT_M}) into a sum of the products of left-moving partial amplitudes and right-moving partial amplitude, where the number of terms is given in \cite{KLT},
\begin{eqnarray}
&(M-3)!&\big[\frac{1}{2}(M-3)\big]!\big[\frac{1}{2}(M-3)\big]!, \ \text{if M is odd}\\
&(M-3)!&\big[\frac{1}{2}(M-4)\big]!\big[\frac{1}{2}(M-2)\big]!, \ \text{if M is even}.
\end{eqnarray}
There is a large number of the channels involved in this sum: since each partial amplitude contains $2^{M-2}(2M-5)!!/(M-1)!$ channels
a naive counting suggests that there are  
\begin{equation}
\frac{2^{2M-4}}{(M-1)!(M-1)(M-2)}\bigg[(\frac{1}{2}(M-3))! (2M-5)!!\bigg]^2
\end{equation}
$c_i n_j$ terms in the Yang-Mills amplitude if $M$ is odd and a similarly large number if $M$ is even.  However, the zero slope limit for gluon sector is the same as Yang-Mills theory, so only the ``diagonal" terms survive, 
\begin{equation}
\mathcal A_{\text{M-gluon}}^{\text{het}}|_{\alpha ' \to 0}=\mathcal A^{\text{YM}}_M=\sum_{P} \frac{c_{\alpha,P}n_{\alpha,P}}{P}.
\end{equation} 
As in the 4-point case, we conjecture that this simplification uses all the color identities for $c_j$'s and all the independent kinematic identities for $n_j$'s. This is explicitly checked for $M=5$, to which we now turn.

%As the four-point case, we can first simplify the expression (\ref{KLT_M}) by contour integral and the result terms would have the form $k^{2M-6} A_\alpha \bar A_{\alpha '}$. Then, using the kinematic identities, the color identities and kinetic energy-momentum relations, we can finally eliminate all the cross terms, $c_P n_{P '}/{P''}$, with only the ``diagonal" terms $c_P n_{P}/{P}$ left as Eq.(\ref{KLT_M}).

\section{The 5-gluon Tree Amplitude Example}

Let us now illustrate the above discussion with the 5-point example. The properties of the 5-point open string amplitudes and their zero slope limit are discussed in Ref.\cite{BDV,Mafra}, while the Yang-Mills amplitudes are discussed in Ref.\cite{BCJ}. So here we shall emphasize the heterotic string properties. We shall also discuss the gauge choice issue to get a sense of the underlying structure.
% of the partial amplitudes 
% to find the low energy limit and the Jacobi identities by the contour integral method.

\subsection{Left-moving amplitudes and color identities}

We fix the three points $x_1=0$, $x_4=1$ and $x_5=\infty$, so the amplitude $A_{12345}$ has the integration region
\begin{equation}
\int_0^1 d x_2 \int_{x_2}^1 d x_3
\end{equation}
As in Eq.(\ref{contour_identity_separated}), the contour integral in $x_2$ over the line above the real axis gives
\begin{equation}
0=\mathbf A_{12345}^{L(c)}+\mathbf A_{23145}^{L(c)} e^{i\frac{\alpha '}{2}(k_1 \cdot k_2 +k_1 \cdot k_3)}+\mathbf A_{21345}^{L(c)} e^{i\frac{\alpha '}{2}(k_1 \cdot k_2)} -\mathbf A_{14325}^{L(c)}  e^{i\frac{\alpha '}{2}(k_1 \cdot k_5)}
\label{contour_identity_5}
\end{equation}
where in this case, $\alpha=1$, $\beta=2$, $\gamma=3$ and $\delta=34$. The correspond poles are $p_A=1$, $p_B=1$, $p_C=1$ and $p_D=-(k_4+k_5)^2$. 

Here we follow the conventions in Ref.\cite{BCJ} in labeling the color factors within the left-moving amplitudes, in the limit $\alpha '\to 0$, 
%\begin{eqnarray}
%A_{12345}^{L(c)}&=&\frac{c_4}{s_{23}s_{45}}+\frac{c_5}{s_{34}s_{15}}+\frac{c_1}{s_{12}s_{45}}+\frac{c_2}{s_{23}s_{15}}+\frac{c_3}{s_{12}s_{34}}\nonumber \\
%A_{23145}^{L(c)}&=&\frac{-c_4}{s_{23}s_{45}}+\frac{c_{15}}{s_{13}s_{45}}+\frac{c_7}{s_{14}s_{23}}+\frac{c_9}{s_{13}s_{25}}+\frac{c_6}{s_{25}s_{14}}\nonumber\\
%A_{21345}^{L(c)}&=&\frac{c_8}{s_{25}s_{34}}+\frac{-c_{15}}{s_{13}s_{45}}+\frac{-c_1}{s_{12}s_{45}}+\frac{-c_9}{s_{25}s_{13}}+\frac{-c_3}{s_{12}s_{34}}\nonumber \\
%A_{14325}^{L(c)}&=&\frac{c_8}{s_{34}s_{25}}+\frac{c_5}{s_{34}s_{15}}+\frac{c_6}{s_{14}s_{25}}+\frac{c_2}{s_{23}s_{15}}+\frac{c_7}{s_{14}s_{23}}
%\label{sub_amplitude_color_5}
%\end{eqnarray}
\begin{eqnarray}
&&  A^{L(c)}_5(1,2,3,4,5) \equiv {c_1\over s_{12}s_{45}}+{c_2\over s_{23}s_{51}}
+{c_3\over s_{34}s_{12}}+{c_4\over s_{45}s_{23}}+{c_5\over s_{51}s_{34}}\,, \nn\\
&&  A^{L(c)}_5(1,4,3,2,5) 
         \equiv {c_{6}\over s_{14}s_{25}}+{c_5\over s_{43}s_{51}}
+{c_7\over s_{32}s_{14}}+{c_{8}\over s_{25}s_{43}}
+{c_2\over s_{51}s_{32}} \,, \nn\\
&&  A^{L(c)}_5(1,3,4,2,5) \equiv {c_{9}\over s_{13}s_{25}}-{c_5\over s_{34}s_{51}}
+{c_{10}\over s_{42}s_{13}}-{c_{8}\over s_{25}s_{34}}
+{c_{11}\over s_{51}s_{42}}\,, \nn\\
&&  A^{L(c)}_5(1,2,4,3,5) \equiv {c_{12}\over s_{12}s_{35}}
+{c_{11}\over s_{24}s_{51}}-{c_3\over s_{43}s_{12}}
+{c_{13}\over s_{35}s_{24}}-{c_5\over s_{51}s_{43}}, \nn\\
&&  A^{L(c)}_5(1,4,2,3,5) \equiv {c_{14}\over s_{14}s_{35}}-
{c_{11}\over s_{42}s_{51}}-{c_7\over s_{23}s_{14}}-
{c_{13}\over s_{35}s_{42}}-{c_2\over s_{51}s_{23}} \,, \nn\\
&&  A^{L(c)}_5(1,3,2,4,5) \equiv {c_{15}\over s_{13}s_{45}}-
{c_2\over s_{32}s_{51}}-{c_{10}\over s_{24}s_{13}}-{c_4\over s_{45}s_{32}}-
{c_{11}\over s_{51}s_{24}} \,,
\label{sub_amplitude_color_5}
\end{eqnarray}
where $s_{ij} = s_{ji}=-(k_i+k_j)^2$. The rest left-moving partial amplitudes are not linear independent, so they do not contain new $c$'s, for example, 
\begin{equation}
A_{21345}^{L(c)}=\frac{c_8}{s_{25}s_{34}}+\frac{-c_{15}}{s_{13}s_{45}}+\frac{-c_1}{s_{12}s_{45}}+\frac{-c_9}{s_{25}s_{13}}+\frac{-c_3}{s_{12}s_{34}}.
\end{equation}

% so the order in the pole is not relevant, say, $(45)=(54), (12)(45)=(45)(12)$.  
%Here we follow the conventions in Ref.\cite{BCJ} in labeling the color factors. 
%To simplify the notation, rather than using the notation, $c_{\alpha, p}$, we just define one of  $c_{\alpha, p}$ to be $c_j$, where the index $j$ is choose in the convention of \cite{BCJ} so we can compare our result with theirs. 

The leading real part of the open string identity (\ref{contour_identity_5}) yields 
\begin{equation}
0=A_{12345}^{L(c)}+A_{23145}^{L(c)} +A_{21345}^{L(c)} -A_{14325} ^{L(c)} 
%\label{contour_identity_5}
\end{equation}
which determines the relative sign for the $c$'s appearing in different sub-amplitudes. Eq.(\ref{sub_amplitude_color_5}) has already taken advantage of this identity.  

The leading order of the imaginary part of  the identity (\ref{contour_identity_5}) is,
\begin{equation}
0=A_{23145}^{L(c)} (s_{12} +s_{13})+A_{21345}^{L(c)} s_{12} +A_{14325} ^{L(c)} s_{15}
\end{equation}
This identity contains $20$ terms, each of which may contain a ``single pole", like ${c_{15}}/{s_{45}}$ or a ``double pole", like ${-s_{12}c_4}/{s_{23}s_{45}}$. The basic strategy to simplify the identity is to combine the terms with the same $c_i$ together and use the kinetic identities like,
\begin{equation}
s_{12}+s_{13}+s_{23}=s_{45}
\end{equation}
to eliminate the double poles. Now each $c_i$ is just the residue of a single pole. We are left with $3$ single-pole terms for each of 
the $4$ different single poles: $s_{45}$, $s_{23}$, $s_{25}$ and $s_{34}$. 
\begin{eqnarray}
\frac{-c_6+c_8+c_9}{s_{25}}+\frac{-c_3+c_5-c_8}{s_{34}}- \frac{c_1-c_4-c_{15}}{s_{45}}- \frac{-c_2+c_4+c_7}{s_{23}}=0.
\label{Ropen51c}
\end{eqnarray}
Since the coefficient of each pole term must vanish, we get four color identities, 
\begin{eqnarray}
c_4+c_{15}-c_1&=&0 \nonumber \\
c_4+c_7-c_2&=&0 \nonumber \\
c_8+c_9-c_6&=&0 \nonumber \\
c_3+c_8-c_5&=&0
\label{i5c1}
\end{eqnarray}
which correspond to residue of the four single poles $s_{45}$, $s_{23}$, $s_{25}$ and $s_{34}$, respectively. Note that the first identity is the special case of the general identity (\ref{Jacobi_string}) with $\alpha=1$, $\beta=2$, $\gamma=3$ and $\delta=34$. 

We can repeat the above contour integral argument to get the rest of the color identities. Similarly, for the configuration $x_5=0$, $x_3=1$, $x_4=\infty$ and $\int_0^1 d x_1 \int_{x_1}^1 d x_2$, the contour integral in $x_1$ gives,
\begin{eqnarray}
\label{i5c2}
c_6-c_7-c_{14}&=&0\nn\\
-c_2+c_4+c_7&=&0\nn\\
-c_3+c_5-c_8&=&0\nn \\
c_3+c_{12}-c_1&=&0
\end{eqnarray}
For the configuration $x_2=0$, $x_4=1$, $x_5=\infty$ and $\int_0^1 d x_1 \int_{x_1}^1 d x_3$, the contour integral in $x_1$ gives,  
 \begin{eqnarray}
 \label{i5c3}
-c_{9}+c_{10}+c_{15}&=&0\nn \\
c_1-c_4-c_{15}&=&0\nn\\
-c_3+c_5-c_8&=&0\nn \\
c_2-c_5-c_{11}&=&0  .
 \end{eqnarray}
while the configuration $x_3=0$, $x_4=1$, $x_5=\infty$ and $\int_0^1 d x_1 \int_{x_1}^1 d x_2$, the contour integral in $x_1$ gives,
 \begin{eqnarray}
 \label{i5c4}
-c_{10}+c_{11}-c_{13}&=&0\nn \\
c_2-c_5+c_{11}&=&0\nn \\
c_3+c_{12}-c_1&=&0\nn \\
c_1-c_4-c_{15}&=&0.
 \end{eqnarray}

So we get all the 10 color identities from four contour integrals.  Other contour integral considerations do not yield additional identities. Since the sum of any $9$ of the 10 identities yields the remaining one, there are 9 independent color identities. 

%(the identity $c_{3}+c_{8}-c_5=0$ appears $3$ times).

Note that these color identities can be read off directly by looking at the diagrams. Consider any 4 (internal and/or external) lines that are connected by another internal line. These 4 lines can be connected 3 different ways, the equivalent of  $"s, t, u"$ channels. The corresponding 3 color factors obey a color identity, which is simply the Jacobi identity multiplied by a common factor of the structure constant. Actually, this is a more efficient way to find the color identities.

\subsection{Right-moving amplitude and the gauge choices}

Now turn to the right-moving part for the gluon momenta and polarizations. The analysis is the same as that for the color factors except for the crucial issue of the contact contributions or the gauge choice. Let us define the $n_i$'s to be local, so that,
\begin{eqnarray}
&&  A^{tree}_5(1,2,3,4,5) \equiv {n_1\over s_{12}s_{45}}+{n_2\over s_{23}s_{51}}
+{n_3\over s_{34}s_{12}}+{n_4\over s_{45}s_{23}}+{n_5\over s_{51}s_{34}}\,, \nn\\
&&  A^{tree}_5(1,4,3,2,5) 
         \equiv {n_{6}\over s_{14}s_{25}}+{n_5\over s_{43}s_{51}}
+{n_7\over s_{32}s_{14}}+{n_{8}\over s_{25}s_{43}}
+{n_2\over s_{51}s_{32}} \,, \nn\\
&&  A^{tree}_5(1,3,4,2,5) \equiv {n_{9}\over s_{13}s_{25}}-{n_5\over s_{34}s_{51}}
+{n_{10}\over s_{42}s_{13}}-{n_{8}\over s_{25}s_{34}}
+{n_{11}\over s_{51}s_{42}}\,, \nn\\
&&  A^{tree}_5(1,2,4,3,5) \equiv {n_{12}\over s_{12}s_{35}}
+{n_{11}\over s_{24}s_{51}}-{n_3\over s_{43}s_{12}}
+{n_{13}\over s_{35}s_{24}}-{n_5\over s_{51}s_{43}}, \nn\\
&&  A^{tree}_5(1,4,2,3,5) \equiv {n_{14}\over s_{14}s_{35}}-
{n_{11}\over s_{42}s_{51}}-{n_7\over s_{23}s_{14}}-
{n_{13}\over s_{35}s_{42}}-{n_2\over s_{51}s_{23}} \,, \nn\\
&&  A^{tree}_5(1,3,2,4,5) \equiv {n_{15}\over s_{13}s_{45}}-
{n_2\over s_{32}s_{51}}-{n_{10}\over s_{24}s_{13}}-{n_4\over s_{45}s_{32}}-
{n_{11}\over s_{51}s_{24}} \,,
\label{Indep5}
\end{eqnarray}

The contour integral argument is exactly the same after we replace, in the partial amplitudes (\ref{sub_amplitude_color_5}), 
\begin{equation}
L\to R, \ (c)\to (v), \ c_j \to n_j 
\end{equation}
For example, now the contour integral in $x_2$ for $\mathbf A_{12345}^{R(v)}$, analogous to the relation (\ref{contour_identity_5}), reads
\begin{equation}
0=\mathbf A_{12345}^{R(v)}+\mathbf A_{23145}^{R(v)} e^{i\frac{\alpha '}{2}(k_1 \cdot k_2 +k_1 \cdot k_3)}+\mathbf A_{21345}^{R(v)} e^{i\frac{\alpha '}{2}(k_1 \cdot k_2)} -\mathbf A_{14325}^{R(v)}  e^{i\frac{\alpha '}{2}(k_1 \cdot k_5)}
\end{equation}
and its imaginary part in the zero slope limit yields,
\begin{equation}
\label{Ropen51n}
%\frac{-n_{9}+n_{10}-n_{15}}{s_{13}}+\frac{n_1-n_4-n_{15}}{s_{45}}- %\frac{-n_3+n_5-n_8}{s_{34}}-\frac{n_2-n_5+n_{11}}{s_{15}}=0 
\frac{-n_6+n_8+n_9}{s_{25}}+\frac{-n_3+n_5-n_8}{s_{34}}- \frac{n_1-n_4-n_{15}}{s_{45}}- \frac{-n_2+n_4+n_7}{s_{23}}=0.
%\frac{n_4+n_{15} -n_1}{s_{45}} + \frac{n_4+n_{7} -n_2}{s_{23}}  +\frac{n_8+n_{9} -n_6}{s_{25}}  +\frac{n_3+n_{8} -n_5}{s_{34}}  = 0
\end{equation}
The residue of each pole term must vanish. However, each of the 4 terms do not have to vanish by itself. That is, the non-pole terms (the contact terms) can cancel among the 4 terms. In particular, $n_3+n_{8} -n_5 = \Delta (k_i, \zeta_i) s_{34}$. 
Consider the gauge transformation  
\begin{equation}
\label{AA3}
n_3 \to n'_3=n_3 + \beta s_{12}
\end{equation}
where the kinematic function $\beta (k_i, \zeta_i)$ is local.
Then invariance of 
$$A_{12345}^{R(v)}=\frac{n_4}{s_{23}s_{45}}+\frac{n_5}{s_{34}s_{15}}+\frac{n_1}{s_{12}s_{45}}+\frac{n_2}{s_{23}s_{15}}+\frac{n_3}{s_{12}s_{34}}$$
implies that
\begin{equation}
\label{BB5}
n_5 \to n'_5=n_5 - \beta s_{15}
\end{equation}
and invariance of all the remaining partial amplitudes means
\begin{equation}
\label{CC8}
n_8 \to n'_8=n_8 + \beta s_{25}
\end{equation}
It follows that $A^{\text{YM}}_M$ is invariant under this triplet of simultaneous transformations.
Under this gauge transformation, we can choose $\beta$ such that
\begin{equation}
\label{E}
n'_3+n'_{8} -n'_5 = \Delta s_{34} + \beta (s_{12} + s_{15} + s_{25}) = (\Delta + \beta) s_{34} =0
\end{equation}
Now we can repeat this process for  $n_4+n_{15} -n_1$ and $n_4+n_{7} -n_2$ to obtain
\begin{equation}
n'_4+n'_{15} -n'_1=0, \quad \quad n'_4+n'_{7} -n'_2=0
\end{equation}
It then follows that $n'_8+n_{9} -n_6=0$. That is, we need to make 3 simultaneous gauge transformations to obtain the above 4 kinematic identities from the string identity (\ref{Ropen51n}).
The key is that the gauge transformation always involves 3 $n_j$ at a time. It is precisely such a triplet of $n_j$ that appears in each kinematic identity. 

With this preliminary discussion, we are now ready to consider the full set. Since there are $(M-3)!=2$ basis amplitudes out of $(M-2)!=6$ amplitudes $A^{tree}$s, there are 4 independent relations among them. We already obtained one in (\ref{Ropen52n}).
%Consider the partial amplitude $A_{12345}$, we fix $x_1=0$, $x_4=1$ and $x_5=\infty$ and integrate over
%\begin{equation}
%\int_0^1 d x_2 \int_{x_2}^1 d x_3...
%\end{equation}
%The contour integral in $x_2$ over $(-\infty, \infty)$ would give 
%\begin{equation}
%\label{Ropen51n}
% \frac{-n_6+n_8+n_9}{s_{25}}+\frac{-n_3+n_5-n_8}{s_{34}}- \frac{n_1-n_4-n_{15}}{s_{45}}- \frac{-n_2+n_4+n_7}{s_{23}}=0.
%\end{equation}
%which is exactly (\ref{condition_one}). 
Similarly, as we did for the left-moving amplitude, for the configuration $x_5=0$, $x_3=1$, $x_4=\infty$ and $\int_0^1 d x_1 \int_{x_1}^1 d x_2$, the contour integral in $x_1$ gives,
\begin{equation}
\label{Ropen52n}
\frac{n_6-n_7-n_{14}}{s_{14}}+ \frac{-n_2+n_4+n_7}{s_{23}}- \frac{-n_3+n_5-n_8}{s_{34}} -
  \frac{n_3+n_{12}-n_1}{s_{12}}=0
\end{equation}
%which is the first equation in (\ref{condition_rest}). 
For the configuration $x_2=0$, $x_4=1$, $x_5=\infty$ and $\int_0^1 d x_1 \int_{x_1}^1 d x_3$, the contour integral in $x_1$ gives the second equation in (\ref{condition_rest}), 
 \begin{equation}
 \label{Ropen53n}
 \frac{-n_{9}+n_{10}+n_{15}}{s_{13}}+\frac{n_1-n_4-n_{15}}{s_{45}}- \frac{-n_3+n_5-n_8}{s_{34}}-\frac{n_2-n_5-n_{11}}{s_{15}}=0 .
 \end{equation}
while the configuration $x_3=0$, $x_4=1$, $x_5=\infty$ and $\int_0^1 d x_1 \int_{x_1}^1 d x_2$, the contour integral in $x_1$ gives the last equation in (\ref{condition_rest}). 
 \begin{equation}
 \label{Ropen54n}
 \frac{-n_{10}+n_{11}-n_{13}}{s_{24}}+ \frac{n_2-n_5+n_{11}}{s_{15}}-\frac{n_3+n_{12}-n_1}{s_{12}}-\frac{n_1-n_4-n_{15}}{s_{45}}=0.
 \end{equation}
Relations from other contour integral identities are redundant. Note that relations  (\ref{Ropen51n}, \ref{Ropen52n}, \ref{Ropen53n}, \ref{Ropen54n}) are gauge invariant. To avoid the gauge dependence issues, one may choose to consider relations among the gauge-invariant partial amplitudes only, which are equivalent to these relations.

It is clear that the residue of each pole term in these relations (\ref{Ropen51n}, \ref{Ropen52n}, \ref{Ropen53n}, \ref{Ropen54n}) must vanish. 
This yields 10 relations, which are the 10 kinematic identities for the residues of the $n_j$'s. Now we like to show that there exists a gauge choice such that every triplet vanishes completely, so we have the 10 kinematic identities,
\begin{eqnarray}
&& \tilde n_3 - \tilde n_5 + \tilde n_8=0 \,,  \nn\\
&& \tilde n_3 - \tilde n_1 + \tilde n_{12}=0\,,  \nn\\
&& \tilde n_4 - \tilde n_1 + \tilde n_{15}=0\,,  \nn\\
&& \tilde n_4 - \tilde n_2 + \tilde n_7=0\,,  \nn\\
&& \tilde n_5 - \tilde n_2 + \tilde n_{11}=0\,,  \nn\\
&& \tilde n_7 - \tilde n_6 + \tilde n_{14}=0\,,  \nn\\
&& \tilde n_8 - \tilde n_6 + \tilde n_9=0\,,  \nn\\
&& \tilde n_{10} -\tilde  n_9 + \tilde n_{15}=0\,,  \nn\\
&&\tilde  n_{10} -\tilde  n_{11} +\tilde  n_{13}=0\,, \nn\\
&&\tilde  n_{13} -\tilde n_{12} +\tilde n_{14}=0
\label{newnjs}
\end{eqnarray}
Note that  one of these 10 identities is redundant.
%This shows the duality between the kinematic and the color identities. 

Because there are $15$ $n_i$'s inside the $6$ defining amplitudes (\ref{Indep5}) above, there are $9$ degrees of freedom to redefine $n_i$'s without affecting the $A^{tree}$'s, which can be realized as,
\begin{eqnarray}
\tilde n_1=n_1+a_{12} s_{45}-a_{45} s_{12}, & & \tilde n_2=n_2+a_{23} s_{15}-a_{15} s_{23}\nn \\
\tilde n_3=n_3+a_{34} s_{12}-a_{12} s_{34}, & &  \tilde n_4=n_4+a_{45} s_{23}-a_{23} s_{45}\nn \\
\tilde n_5=n_5+a_{15} s_{34}-a_{34} s_{15}, & & \tilde n_6=n_6+a_{25} s_{14}-a_{14} s_{25}\nn \\
\tilde n_7=n_7+a_{14} s_{23}-a_{23} s_{14}, & & \tilde n_8=n_8+a_{34} s_{25}-a_{25} s_{34}\nn \\
\tilde n_9=n_9+a_{13} s_{25}-a_{25} s_{13}, & & \tilde n_{10}=n_{10}+a_{13} s_{24}-a_{24} s_{13}\nn \\
\tilde n_{11}=n_{11}+a_{15} s_{24}-a_{24} s_{15}, & & \tilde n_{12}=n_{12}+a_{35} s_{12}-a_{12} s_{35}\nn \\
\tilde n_{13}=n_{13}+a_{35} s_{12}-a_{12} s_{35}, & & \tilde n_{14}=n_{14}+a_{14} s_{35}-a_{35} s_{14}\nn \\
n_{15}=n_{15}+a_{45} s_{13}-a_{13} s_{45}.& &  
\label{redefinition}
\end{eqnarray}
where $a_{34},a_{12},a_{45},a_{23},a_{15},a_{14},a_{25},a_{13},a_{24},a_{35}$ are arbitrary local functions of $k_j$ and $\zeta_j$. The signs are carefully chosen such that the partial amplitude is invariant. Although the number of $a_{ij}$ is $10$, a particular choice,
\begin{eqnarray}
&&(a_{34},a_{12},a_{45},a_{23},a_{15},a_{14},a_{25},a_{13},a_{24},a_{35})\nonumber \\
&=&(s_{34},s_{12},s_{45},s_{23},s_{15},s_{14},s_{25},s_{13},s_{24},s_{35})
\label{redunre}
\end{eqnarray}
does not change any $n_i$, so it is  a trivial redefinition. Therefore we can simply set any one of them to zero. Let us say $a_{35}=0$, so we end up with $9$ degrees of gauge freedom.
We already see that the non-contact terms inside $n_i$ satisfy the dual identity $(n_i+n_j+n_k)|_{\text{residue}}=0$, if the color factors with the same indices satisfy $c_i+c_j+c_k=0$. We like to show that by using a proper redefinition of the $n_i$'s, the dual identities (\ref{newnjs}) hold exactly.

Since the redefinition of the $n_j$'s are realized by the $a$'s in Eq.(\ref{redefinition}), and we like to see whether a choice of the $a$'s exists for the set of kinematic identities (\ref{newnjs}) to hold, these equations can be understood as the equations for the $a$'s. For example, $\tilde n_3 - \tilde n_5 + \tilde n_8=0$ reads, 
\begin{equation}
a_{34}-a_{12}-a_{15}-a_{25}=\frac{-n_3+n_5-n_8}{s_{34}}
\end{equation}
where we used $s_{15}+s_{25}+s_{12}=s_{34}$. Similarly, we can write all the 9 equations in the matrix form, 
\begin{equation}
K a=b
\label{linear}
\end{equation}
where 
\begin{equation}
K=\left(
\begin{array}{ccccccccc}
 1 & -1 & 0 & 0 & -1 & 0 & -1 & 0 & 0 \\
 -1 & 1 & -1 & 0 & 0 & 0 & 0 & 0 & 0 \\
 0 & -1 & 1 & -1 & 0 & 0 & 0 & -1 & 0 \\
 0 & 0 & -1 & 1 & -1 & -1 & 0 & 0 & 0 \\
 -1 & 0 & 0 & -1 & 1 & 0 & 0 & 0 & -1 \\
 0 & 0 & 0 & -1 & 0 & 1 & -1 & 0 & 0 \\
 -1 & 0 & 0 & 0 & 0 & -1 & 1 & -1 & 0 \\
 0 & 0 & -1 & 0 & 0 & 0 & -1 & 1 & -1 \\
 0 & 0 & 0 & 0 & -1 & 0 & 0 & -1 & 1
\end{array}
\right),\ a= \left(
\begin{array}{c}
 a_{34}  \\
  a_{12} \\
  a_{45} \\
  a_{23}  \\
 a_{15} \\
 a_{14} \\
  a_{25} \\
  a_{13}  \\
 a_{24}
\end{array}
\right)
\end{equation}
and
\begin{eqnarray}
&b&=
\bigg (\frac{-n_3+n_5-n_8}{s_{34}} ,
  \frac{n_3+n_{12}-n_1}{s_{12}},
 \frac{n_1-n_4-n_{15}}{s_{45}},
  \frac{-n_2+n_4+n_7}{s_{23}}  ,
 \frac{n_2-n_5-n_{11}}{s_{15}}, \nonumber\\&&
 \frac{n_6-n_7-n_{14}}{s_{14}} ,
  \frac{-n_6+n_8+n_9}{s_{25}},
\frac{-n_{9}+n_{10}+n_{15}}{s_{13}},
   \frac{-n_{10}+n_{11}-n_{13}}{s_{24}}\bigg)^\text{T}
\end{eqnarray}
Here we are trying to find a solution to the $a$'s so that Eq.(\ref{newnjs}) holds.
Given an arbitrary $b$, a solution is always guaranteed if the rank of $K$ equals its size (which is $9$).
However, here $K$ is a degenerate matrix, with  ${\text rank} \ K =5 $, that is, ${\text rank} \ K < 9$.
%{\bf Note that the solution of (\ref{linear}) may not exist, because $M$ is a degenerate matrix, $\text{rank} M=5<9$.} 
So a solution of the $a$'s exists only if 4 (=9-5) constraints among the components of the column vector $b$ are satisfied.
That is, only 5 equations are independent and they generate the remaining 4 equations. For example, 
\begin{eqnarray}
0&=&(a_{25}-a_{14}-a_{34}-a_{13})+(a_{34}-a_{12}-a_{15}-a_{25})\nonumber\\
&&-(a_{45}-a_{12}-a_{13}-a_{23})
-(a_{23}-a_{15}-a_{45}-a_{14})\nonumber\\
&=& \frac{-n_6+n_8+n_9}{s_{25}}+\frac{-n_3+n_5-n_8}{s_{34}}- \frac{n_1-n_4-n_{15}}{s_{45}}- \frac{-n_2+n_4+n_7}{s_{23}}\nonumber\\
\label{condition_one}
\end{eqnarray}
which is a constraint on the original $n_j$'s. Similarly,
\begin{eqnarray}
 \frac{n_6-n_7-n_{14}}{s_{14}}+ \frac{-n_2+n_4+n_7}{s_{23}}- \frac{-n_3+n_5-n_8}{s_{34}} -
  \frac{n_3+n_{12}-n_1}{s_{12}}=0, \nonumber\\
\frac{-n_{9}+n_{10}+n_{15}}{s_{13}}+\frac{n_1-n_4-n_{15}}{s_{45}}- \frac{-n_3+n_5-n_8}{s_{34}}-\frac{n_2-n_5-n_{11}}{s_{15}}=0 ,\nonumber\\
\frac{-n_{10}+n_{11}-n_{13}}{s_{24}}+ \frac{n_2-n_5+n_{11}}{s_{15}}-\frac{n_3+n_{12}-n_1}{s_{12}}-\frac{n_1-n_4-n_{15}}{s_{45}}=0.
\label{condition_rest}
\end{eqnarray}  
These 4 constraints (\ref{condition_one},\ref{condition_rest}) form the necessary and sufficient condition for the existence of a solution  to Eq.(\ref{linear}). Naively, just from the Feynman diagram viewpoint, it is not clear why these conditions hold. However, we see that the open string amplitude identities (\ref{Ropen51n}, \ref{Ropen52n}, \ref{Ropen53n}, \ref{Ropen54n}) yield precisely these 4 relations.
Hence open string identities ensure that (\ref{linear}) has solutions. That is, there exists a gauge choice such that Eq.(\ref{newnjs}) is realized.

Since the column vector $b$ has no pole (i.e., local), because $-n_3+n_5-n_8 \propto s_{34}$ etc., the solution yields a set of local $a$'s. Because the solution for (\ref{linear}) exists, there are $9-\text{rank} \ K=4$ remaining transformations which keep the kinematic identities invariant. They take the forms
\begin{equation}
\left(
\begin{array}{c}
 a_{34}  \\
  a_{12} \\
  a_{45} \\
  a_{23}  \\
 a_{15} \\
 a_{14} \\
  a_{25} \\
  a_{13}  \\
 a_{24}
\end{array}
\right) = f_1
\left(
\begin{array}{c}
 0  \\
  -1 \\
  -1 \\
  0  \\
 1 \\
 0 \\
  0 \\
  0  \\
 1
\end{array}
\right),\  f_2
\left(
\begin{array}{c}
 -1  \\
  0 \\
  1 \\
  0  \\
 -1 \\
 0 \\
  0 \\
  1  \\
 0
\end{array}
\right), \ f_3
\left(
\begin{array}{c}
 1  \\
  0 \\
  -1 \\
  -1  \\
 0 \\
 0 \\
  1 \\
  0  \\
 0
\end{array}
\right), \ f_4
\left(
\begin{array}{c}
 -1  \\
  -1 \\
  0 \\
  1  \\
 0 \\
 1 \\
  0 \\
  0  \\
 0
\end{array}
\right)
\end{equation} 
where $f_i$ are arbitrary local functions of the kinematic variables.
This concludes our proof that,  for the $(M=5)$-point case, the open string identities ensure that there exists choices of $n_i$'s such that all the dual identities $n_i+n_j+n_k=0$ (\ref{newnjs}) hold.
% and the dimension of the ``choice space'' is $4$.

This pattern generalizes to the general $M$-point amplitudes. There are $(2 M-5)!!$ $n_i$'s which appeared in $(M-2)!$ partial amplitudes so there are  $(2 M-5)!!-(M-2)!$ degrees of freedom to redefine the $n_i$'s. We can use $\frac{(M-3)(2 M-5)!!}{3}$ parameters $a_k$ modulo some trivial ones to realization the redefinitions. The number of the effective $a_k$'s should be $(2 M-5)!!-(M-2)!$, which are constraint by a linear equation like (\ref{linear}) if we want to get the kinematic identities. Again the matrix in the linear equation is degenerate with rank $(2 M-5)!!-(2M-5)((M-3)!)$. The open string identities would ensure this linear equation has solutions. So the $n_i$ choice for which the kinematic identities hold exist, and the ``choice space" has the dimension $(M-2)!-(M-3)!=(M-3)!(M-3)$.

\subsection{KLT relation and the $5$-point amplitudes}

We can now use the KLT relation in the zero slope limit to get the 5-point Yang-Mills tree amplitude. Since the heterotic amplitude is a sum over the product of a left-mover and a right mover, and since there are 2 independent partial amplitudes for $A^L$ and 2 for $A^R$, we can express the full amplitude as a sum over $2 \times 2=4$ terms. As shown in Ref.\cite{KLT}, a judicious choice of basis amplitudes allows us to reduce the sum to only 2 terms. There are many equivalent ways to express the 5-point amplitude. For example, 
\begin{eqnarray}
\mathcal A^{\text{het}}_{\text{5-gluon}}(0) &=&g^3 s_{12} s_{34}A_{12345}^{L(c)} A_{21435}^{R(v)}+g^3 s_{13} s_{24} A_{13245}^{L(c)}A_{31425}^{R(v)} \nonumber \\
&=& g^3\bigg(\frac{c_4}{s_{23}s_{45}}+\frac{c_5}{s_{34}s_{15}}+\frac{c_1}{s_{12}s_{45}}+\frac{c_2}{s_{23}s_{15}}+\frac{c_3}{s_{12}s_{34}} \nonumber\bigg) \\ &\times& s_{12} s_{34} \bigg(\frac{-n_{12}}{s_{12}s_{35}}+\frac{-n_6}{s_{14}s_{25}}+\frac{n_3}{s_{12}s_{34}}+\frac{-n_{14}}{s_{14}s_{35}}+\frac{-n_{8}}{s_{34}s_{25}} \bigg) \nonumber \\
&+&g^3\bigg(\frac{c_{15}}{s_{13}s_{45}}+\frac{-c_2}{s_{23}s_{15}}+\frac{-c_{10}}{s_{13}s_{24}}+\frac{-c_{4}}{s_{45}s_{23}}+\frac{-c_{11}}{s_{15}s_{24}}\nonumber\bigg) \\ &\times&  s_{13} s_{24} \bigg(\frac{-n_{9}}{s_{13}s_{25}}+\frac{-n_{14}}{s_{14}s_{35}}+\frac{-n_{10}}{s_{13}s_{24}}+\frac{-n_{6}}{s_{14}s_{25}}+\frac{n_{13}}{s_{24}s_{35}} \bigg) 
\label{KLT_5pts}
\end{eqnarray}
First we go to the gauge choice where all the 9 independent kinematic identities hold.
Then using all 18 of the 9 independent color identities (\ref{i5c1}, \ref{i5c2}, \ref{i5c3},\ref{i5c4}) and the 9 independent kinematic identities (\ref{newnjs}), it is straightforward (but tedious) to show that the 5-point Yang-Mills tree amplitude is reproduced,
\begin{eqnarray}
\mathcal A^{\text{het}}_{\text{5-gluon}}(0)&=& \mathcal A^{\text{YM}}_5=g^3 \sum_{j=1}^{15} \frac{c_j n_j}{P_j} \nonumber \\
&=& g^3 \left(\frac{c_1 n_1}{s_{12}s_{45}} + \frac{c_2 n_2}{s_{23}s_{15}} 
+ \frac{c_3 n_3}{s_{12}s_{34}}  + ... \right) 
\end{eqnarray}
If we want, we can now transform back to the original set of $n_j$ we started with. 
In the choice of the particular way (\ref{KLT_5pts}) to express $\mathcal A^{\text{het}}_{\text{5-gluon}}(0)$, the presence of the diagonal term $c_3 n_3/(s_{12} s_{34})$ is obvious, but the other diagonal terms are not. Choosing a different basis to express $\mathcal A^{\text{het}}_{\text{5-gluon}}(0)$, a different diagonal term will be obvious, but not the rest. It is the presence of the $9+9$ identities that allows us write $\mathcal A^{\text{het}}_{\text{5-gluon}}(0)$ in the diagonal form that is given in $\mathcal A^{\text{YM}}_5$ (\ref{decompositionG}). On the other hand, knowing that $\mathcal A^{\text{het}}_{\text{5-gluon}}(0)=\mathcal A^{\text{YM}}_5$, we can obtain the $9+9$ identities as well, by exploiting the many different but equivalent ways to express $\mathcal A^{\text{het}}_{\text{5-gluon}}(0)$. 

On the other hand, instead of using the dual-Jacobi identities (\ref{newnjs}), one can show Eq.(\ref{KLT_5pts}) is equivalent to the diagonal form (\ref{decompositionG}) 
%the diagonal form $\sum c_i n_i/P_i$ by 
using only the gauge-invariant im-SID's (\ref{Ropen51n}, \ref{Ropen52n}, \ref{Ropen53n}, \ref{Ropen54n}). First, we rewrite all the color factors $c_i$'s in (\ref{KLT_5pts}) in terms of $(M-2)!=6$ of them, say, $c_1$, $c_6$, $c_9$, $c_{12}$, $c_{14}$ and $c_{15}$,
% as the basis, and by this basis,
\begin{eqnarray}
\mathcal A^{\text{het}}_{\text{5-gluon}}(0)&=&c_1 A^{R(v)}_{12345} + c_{12} \bigg(-\frac{s_{34}}{s_{35}} A^{R(v)}_{12345}-A^{R(v)}_{12345}+\frac{s_{13}}{s_{35}}A^{R(v)}_{13245}\bigg)\nonumber \\
&+& c_6\bigg(-\frac{s_{12}s_{34}}{s_{14}s_{25}}A^{R(v)}_{12345}-\frac{s_{12}}{s_{25}}A^{R(v)}_{12345}-\frac{s_{13}s_{34}}{s_{14}s_{25}}A^{R(v)}_{13245}\bigg)+...
\end{eqnarray}
%The coefficient of each $c_i$ seems complicated. However, by 
Using all the im-SID's,
\begin{eqnarray}
  -s_{34} A^{R(v)}_{12345}-s_{35} A^{R(v)}_{12345}+ s_{13} A^{R(v)}_{13245}=s_{35} A^{R(v)}_{12435}, ...
\end{eqnarray}
to simply the expression so each coefficient contains only one $A^{R(v)}$,
\begin{eqnarray}
  \mathcal A^{\text{het}}_{\text{5-gluon}}(0)&=&c_1 A^{R(v)}_{12345}+c_{15} A^{R(v)}_{13245}+c_{12} A^{R(v)}_{12435} \nonumber\\
&&+ c_9 A^{R(v)}_{13425}+c_{14} A^{R(v)}_{14235}+c_6 A^{R(v)}_{13245} \nonumber \\
&=& \sum_i \frac{c_i n_i}{P_i},
\end{eqnarray}
where the last equality is proved in \cite{DelDuca:1999rs} by using the color Jacobi identities only. So the diagonal form (\ref{decompositionG}) is obtained, if only one set of the numerators ($c_i$ here) satisfy the Jacobi identities while the $A^{R(v)}$s satisfy the im-SID. This property should extend to all $M$. The KLT relation simply expresses $\mathcal A^{\text{YM}}_M$ (\ref{decompositionG}) in terms of the $(M-3)!$ basis amplitudes $A^{R(v)}$s. Furthermore, for the 5-graviton tree amplitude, in the same manner, we have two sets of numerators, $n_i$ and $\tilde n_i$. As long as the $n_i$'s satisfy the dual Jacobi identities, the tree amplitude is simplified to the diagonal form,
\begin{eqnarray}
  \mathcal A_5^{grav}=\sum_i \frac{n_i \tilde n_i}{P_i},
\end{eqnarray}
where the other set $\tilde n_i$ need not to satisfy the dual Jacobi identities.

%  that channels like are in the product but the existence of the other channels are not explicit. However, if we use the Jacobi identities, dual BCJ identities, and the kinematic identities, after long but straightforward calculation,

%In deriving this ``diagonal" form, we used all the $9$ Jacobi identities and also the $9$ dual BCJ identities. Hence in this sense, the equality of the KLT result and the field theory result indicate that the Jacobi identities and the dual BCJ identities should hold, as we see in the introduction for the 4-point case.  

%There are $15$ channels for five-point amplitude so there are $15$ $c_i$'s when the relation (\ref{common_channel}) is used. For each channel we have $2$ ways to choose an internal line to get the sub-diagrams. Each Jacobi identity contains $3$ sub-diagrams, hence there are $15\times 2/3=10$ Jacobi identities. However, these identities are not independent--- only $9$ of them are. So we have $15-9=6$ independent $c_i$'s.

\section{Discussion and Remarks}

%Graviton and other scattering amplitude}

The heterotic string also contains the graviton sector, as $\text{(vector)} \otimes \text{(vector)}$.
So we can calculate the graviton scattering amplitudes in the heterotic string theory and then take the limit $\alpha '\to 0$ to get the graviton scattering amplitude in Einstein gravity. The graviton sector has both left and right-moving non-compact momenta, and the closed string amplitude can be separated into the product of left and right-moving open string amplitude, both of which are calculated in the Yang-Mill amplitude. As explained already, the $M$-graviton tree scattering amplitude is 
\begin{equation}
A_{\text{M-graviton}}=\sum_j\frac{\tilde n_j  n_j}{P_j}
\end{equation} 
where $n_j(k_i, \zeta_i)$ contains the polarizations $\zeta^\mu_i$ while $\tilde n_j$ contains the polarization $\xi^{\mu}_i$. Otherwise, they have identical functional forms, i.e., $\tilde n_j (k_i, \xi_i)=n_j(k_i, \xi_i)$. The graviton polarization $\epsilon_{\mu\nu}$ is the symmetrized traceless product in $\xi_{\mu} \zeta_{\nu}$. 

We can easily incorporate the scattering of fermions since spinors are present in the right-moving superstring part of the heterotic string. 
This has been discussed in Ref.\cite{Dixon:1986qv}. Keeping only the leading order in $\alpha'$, we have %Written in the form  
$\mathcal A = \sum_j {n^L_j n^R_j}/{P_j}$,
where the $n^L_j$ and separately the $n^R_j$ obey the same set of identities as the color factors. Here $n^L_j$ can include both colors and/or vectors while $n^R_j$ can include both vectors and fermions. These include scattering amplitudes involving gravitons, gluons, gravitinos and gluinos. The resulting identities should be very helpful in the evaluation of scattering amplitudes.
 
Until now, we consider general polarizations of  gluons, $\zeta_i$, and of graviton, $\epsilon_i$, where $\zeta_{i,\mu} k_i^\mu =0$ and $\epsilon_{i,\mu\nu}  k_i^\mu =\epsilon_{i\mu\nu}  k_i^\nu =0$. However, we can consider special polarizations to simplify the computation, for example, by using the spinor helicity formalism  \cite{SpinorHelicity} \cite{XuZhangChang} \cite{GunionKunszt}. The polarization of the gluon is chosen to be,
\begin{equation}
\zeta^{\pm}_\mu (k_i,q_i)=\pm \frac{\langle q_i^{\mp} |\gamma_\mu | k^{\mp}_i\rangle}{\sqrt 2 \langle q^{\mp}_i|k^{\mp}_i \rangle }
\end{equation} 
where $q_i$ is the reference momentum. In this convention, by careful choices of the $q_i$, many terms in the Yang-Mills amplitude vanish. For example, within the four-point partial amplitude $A(1^-,2^-,3^+,4^+)$, $n_t=0$. In this case, the kinematic identity $n_s+n_t+n_u=0$ implies  $n_s=-n_u$ and we just need to consider one channel. 
%So the combination of the the kinematic identities and spinor helicity formalism can simplify the computation further.

Spinor helicity formalism is used for the graviton polarization \cite{Spehler:1991yw},
\begin{equation}
\epsilon_{\mu\nu}^{++}(k_i,q_i)=\zeta^+_\mu(k_i,q_i) \zeta^+_\nu(k_i,q_i), \ \epsilon_{\mu\nu}^{--}(k_i,q_i)=\zeta^-_\mu(k_i,q_i) \zeta^-_\nu(k_i,q_i)
\end{equation} 
%so the graviton polarization tensor is explicitly decomposed into two identical polarization vectors and we do not need to distinguish the formal left-moving polarization $\xi$ and the right-moving polarization $\zeta$ in the KLT product. Both the left and right hands give the same $n_i$, which is already calculated in Yang-Mills theory with the help of the spinor helicity formalism and the kinematic identities.  
For example, since the Yang-Mills amplitude with only one gluon with opposite helicity vanishes: $\mathcal A^{\text{YM}}_M(1^-, 2^+, 3^+, ... ,M^+)=0$, the corresponding $\mathcal A^{grav}_M(1^{--}, 2^{++}, 3^{++}, ... ,M^{++})$ also vanishes.

The heterotic string involves modes in higher representations of the Lie group. Keeping them will introduce generalized ``structure constants" $f^{abC}$, $f^{aBC}$ and $f^{ABC}$, where capital letters $A, B, C$ signify modes in higher representations (than the adjoint representation). Then the string amplitude identities (similar to (\ref{contour_identity})) will yield the corresponding {\it generalized 
Jacobi identities} among them.
As shown in Ref.\cite{zhu}, the kinematic identities can be extended to include the scattering of massive particles. Since the open string amplitudes are multivariable-integrals involving the Koba-Nielsen variables, such Koba-Nielsen integrals can be generalized to include massive particles with higher spins and so obtain the kinematic identities in the scattering of massive particles. 

One can also start with $D=10$ dimensions and compactify (toroidally) 6 of them. The resulting theory in the zero slope limit is a low energy effective $\mathcal N=4$ supergravity theory. This allows us to study scattering amplitudes in $\mathcal N=4$ supersymmetric Yang-Mills theory as well as in  $\mathcal N=4$ supergravity theory. The identities in the tree amplitudes will be carried over to the loop diagrams using the unitarity method. This should provide a better understanding of the loop amplitudes in the $\mathcal N=4$ theory.

In this paper, we have restricted our discussions to the scattering of massless particles only. However, the analysis of Ref.\cite{zhu, Goebel:1980es} strongly suggests that the kinematic identities can be generalized to massive particles as well. It is clear that this subject matter is still wide open.

\vspace{0.7cm}

\noindent {\bf Acknowledgments}

\vspace{0.2cm}

This study was initiated in a discussion with Zvi Bern that we gratefully acknowledge. We also thank him for very valuable comments on the manuscript.
This work is supported by the National Science Foundation under grant PHY-0355005.

\vspace{0.6cm}

\appendix

\section{Lie algebra and cocycles}

Here $\alpha '=\frac{1}{2}$ when not explicitly displayed.
To be specific, we shall consider the Lie group $G$ to be $U(N)$ or $SO(2N)$ only.
%a simply-laced Lie group associated with the heterotic string, but need not to be $E_8 \times E_8$ or $SO(32)$ because now we are only considering the tree level string amplitude. 
The Cartan sub-Lie algebra is $\mathsf \ h$, and we denote 
\begin{equation}
p^I, I=1,..., \mbox{dim} \mathsf \ h
\end{equation}  
as its basis. The roots of $G$ are $E_K$, for which, $K$ is a vector on the root lattice,
\begin{equation}
[p^I,E_K]=i K^I E_K.
\end{equation}
$G$ is simply-laced, so all the roots have the same length square,
\begin{equation}
K\cdot K=1/\alpha '=2.
\end{equation}
There are $N(N-1)$ roots in $U(N)$, spanned by the $N$-vectors $K=\pm (..., +1,...,-1,...)$. For $SO(2N)$, we have the additional roots $K=\pm (..., +1,..., +1,...)$.
To get the complete commutation relations, we introduce the cocycle \cite{KLT}: Let $P^I=n_i e^I_i$ and $K^I=m_i e^I_i$ where $e_i$ is the basis of the root lattice, we define
\begin{equation}
P \star K=\sum_{i>j} n_i m_j (e_i \cdot e_j).
\end{equation}
Then we have,
\begin{equation}
 [E_{K_1},E_{K_2}] = 
  \begin{cases} 
  i (-1)^{K_1 \star K_1} K_1^I p^I & \text{if } K_1 \cdot K_2 =-2 \\
  i (-1)^{K_2 \star K_1} E_{K_1+K_2} & \text{if } K_1 \cdot K_2 =-1  \\
   0 & \text{if } K_1 \cdot K_2 \geq 0.
  \end{cases}
\label{commutator}
\end{equation}

The above commutator relations completely determined the structure constants,
\begin{equation}
[T^a,T^b]=i \sqrt 2 {f^{ab}}_c T^c =  \tilde { f^{ab}}_c T^c
\end{equation}
Note that the $E_K$ are in the raising and lowering basis. We must distinguish the upper and lower index since in the $\{p^I, E_K\}$ basis the invariant inner product, $\tr(T^a T^b)\equiv g^{ab}$, in general is not $\delta^{ab}$.  The invariant product means,
%\begin{equation}
%{f_{ab}}^d g_{dc} +{f_{ac}}^d g_{db} =0
%\label{invariant_product}
%\end{equation}
%\begin{equation}
$f^{abc}=f^{cab}$
%\label{cyclic}
%\end{equation}
where we raise and lower the indices by $g^{ab}$ and $g_{ab}$ respectively.

%Because $G$ is a compact Lie group, there is an invariant inner product on its Lie algebra, which equals the trace of any non-trivial representation,
%\begin{equation}
%(T^a,T^b) \propto \tr _r(T^a T^b)
%\end{equation} 
%up to an overall factor $C(r)$. For simplicity, we use the notation $\tr$ for the invariant inner product and determine the normalization by,

Now we determine the $g_{ab}$ in the $\{p^I, E_K\}$ basis. First, we define,
\begin{equation}
\tr(p^I p^J)=\delta_{IJ},\ \tr(p^I E_{K})=0
\label{ppnorm}
\end{equation}
Second, 
Plug in $T_a=E_K$,  $T_b=E_{-K}$  and $T_c=p^I$ into the invariant relation $f^{abc}=f^{cab}$,  it is easy to see that  by Eq.(\ref{commutator})
\begin{equation}
\tr(E_{K} E_{-K})=(-1)^{K\star K}.
\label{KKnorm}
\end{equation}
Third, if $K_1+K_2 \not =0$, 
\begin{equation}
\tr(E_{K_1} E_{K_2})=0.
\label{zeroKKnorm}
\end{equation}
Eq.(\ref{ppnorm}), (\ref{KKnorm}) and (\ref{zeroKKnorm}) completely fixed the $g^{ab}$. The Jacobi identity is
\begin{equation}
{f^{ab}}_e f^{cde}+{f^{ca}}_e f^{bde} +{f^{bc}}_e f^{ade}=0.
\end{equation}

%\begin{equation}
%\tr ([T^a,T^b][T^c,T^d])+\tr ([T^c,T^a][T^b,T^d]) +\tr([T^b,T^c][T^a,T^d])=0.
%\end{equation}

%or equivalently
%\begin{equation}

%\tr ([[T^a,T^b],T^c]T^d)+\tr ([[T^c,T^a],T^b]T^d) +\tr([[T^b,T^c],T^a]T^d)=0.
%\end{equation}

The vertex operator (left-moving part) for a gluon whose color index is not in Cartan sub-Lie algebra is 
\begin{equation}
V(x;k,K)=e^{\frac{i}{2} k_{\mu} x^{\mu}+i K_I x^I} (-1)^{P\star K},
\label{vertex_operator}
\end{equation}
while for a gluon whose color index is in Cartan sub-Lie algebra is,
\begin{equation}
-i \zeta_I \dot X^I e^{\frac{i}{2} k_{\mu} x^{\mu}},
\label{vertex_operator_vector}
\end{equation}
where we used the convention of Ref. \cite{Polchinski:1998rq}. In the latter case, we can formally define $K=0$. 

The string tree amplitude can be viewed as the OPE's expectation value, say,
\begin{equation}
\langle 0 | V(x_1;k_1,K_1) V(x_1;k_2,K_2) ... V(x_n;k_n,K_n)|0 \rangle 
\end{equation}  
where we suppressed the integral over $x_i$. The co-cycle part gives, 
\begin{equation}
co(12...n)\equiv (-1)^{K_1 \star K_1 +\sum_{1<i<j\leq n} K_j \star K_i}
\end{equation} 
It is easy to check that when the discrete momentum is conserved, i.e., $\sum_{i=1}^n K_i=0$, the notation $co(12...n)$ has the following properties:
\begin{itemize}
\item Cyclic permutation. 
\begin{equation}      
co(12...n)=co(n12...n-1)
\end{equation} 
\item Adjacent transpositions. 
\begin{equation}
co(12...ij...n) \cdot (-1)^{K_i \cdot K_j}=co(12...ji...n),
\label{adjacent}
\end{equation}
when $i$ and $j$ is a pair of adjacent indices. 
\end{itemize}

\section{Explicit determination of the color factors for the $4$-point amplitude}

Because Eq.(\ref{KLT_4pts}) gives the correct Yang-Mills 4-gluon scattering amplitude, we know that the factor $c_s$ must contain the color index like $\tilde f^{a_1 a_2 b} \tilde f^{b a_3 a_4 }$. In this appendix, we explicit calculate the $c$'s and hence check Eq.(\ref{c_ff}). The pattern for general $M$ should be clear. 

The emergence of the Lie group $G$ in the heterotic string bosonic construction is interesting. To take advantage of the discrete momenta, we have to distinguish the generators in the Cartan-sub Lie algebra and those corresponding to vectors in the root lattice. The calculation of the color factors $c$'s in the 2 cases are different. However, as expected, the end result puts all the color indices on an equal footing
% when we finished all the cases, it is clear that the result of $c$'s are uniform in the generators, 
as claimed by Eq.(\ref{c_ff}).

\subsection{Four color indices as root vectors}

In this case, all the vertex operators contain $K^I$ but no $\zeta^I$. As usual, the Mandelstam variables are defined
\begin{eqnarray}
S=-(K_1+K_2)^2, U=-(K_1+K_3)^2, T=-(K_1+K_4)^2,
\end{eqnarray}
and $S+T+U=-4/\alpha ' $. For simplicity, we set $\alpha '=1/2$ when it is combined with the discrete momenta but still keep $\alpha '$ when it is multiplying the spacetime momentum.
 
We can write the amplitude in terms of the Beta functions, in the zero slope limit,
\begin{eqnarray}
A_{L,1234}^{(c)}&=&\frac{\alpha '}{4} co(1234)\cdot B(-\frac{\alpha ' s}{4}-\frac{1}{2} S-1,-\frac{\alpha ' t}{4}-\frac{1}{2} T-1)\nonumber \\
&\sim& co(1234)\cdot -\frac{1}{ s}(\delta_{S,-2}+K_1 \cdot K_3 \delta_{S,0}) \nonumber \\
&& +co(1234) \cdot (-1)^{T/2}\frac{1}{ t}(\delta_{T,-2}+K_1 \cdot K_2 \delta_{T,0}).
\end{eqnarray}
Here ``$\sim $'' means the lowest order in energy, i.e., in $\alpha 's$ etc. Comparing with Eq.(\ref{definition_c}), we have, 
\begin{eqnarray}
c_s&=& -co(1234)\cdot (\delta_{S,-2}+K_1 \cdot K_3 \delta_{S,0})\\
c_t&=& -co(1324) \cdot  (\delta_{T,-2}+K_1 \cdot K_2 \delta_{T,0}).
\end{eqnarray}
Similarly, we have
\begin{eqnarray}
A_{L,2134}^{(c)}&=&co(2134)\cdot B(-\frac{\alpha ' s}{4}-\frac{1}{2} S-1,-\frac{\alpha ' u}{4}-\frac{1}{2} U-1)\nonumber \\
&\sim& co(2134)\cdot -\frac{1}{ s}(\delta_{S,-2}-K_1 \cdot K_3 \delta_{S,0}) \nonumber \\
&& +co(2134) \cdot \frac{1}{ u}(\delta_{U,-2}+K_2 \cdot K_3 \delta_{U,0}).
\end{eqnarray}
Comparing with Eq.(\ref{definition_c}) again,  
\begin{eqnarray}
c_s&=& co(2134)\cdot   (\delta_{S,-2}-K_1 \cdot K_3 \delta_{S,0})\\
c_u&=& -co(2134) \cdot  (\delta_{U,-2}+K_2 \cdot K_3 \delta_{U,0}).
\end{eqnarray} 
Again, we get the factor $c_s$. Note that since $co(2134) \cdot (-1)^{S/2}=co(1234)$, the two results are identical. This is a consequence of the contour integral argument. Now we can compare them with the commutators. From the normalization convention in Eq.(\ref{commutator}), (\ref{ppnorm}) and  (\ref{KKnorm}),
\begin{eqnarray}
\tilde f^{a_1 a_2 b} \tilde f^{c a_3 a_4 } g_{b c} &=&-co(1234)\cdot (\delta_{S,-2}+K_1 \cdot K_3 \delta_{S,0})\\
\tilde {f^{a_3 a_1 b}} \tilde f^{c a_2 a_4 } g_{b c} &=&-co(1324)\cdot (\delta_{T,-2}+K_1 \cdot K_2 \delta_{T,0})\\
\tilde {f^{a_2 a_3 b}} \tilde f^{c a_1 a_4 } g_{b c} &=&-co(2134)\cdot (\delta_{U,-2}+K_2 \cdot K_3 \delta_{U,0}).
\end{eqnarray}
Hence in this case, Eq.(\ref{c_ff}) is checked explicitly.

\subsection{Three color indices as root vectors}

Without loss of generality, we set the first vertex operator has to have its color index in the Cartan subalgebra, i.e., $K_1=0$. The string amplitude calculation is straightforward; here we just keep the lowest order in $\alpha ' s$ etc,
\begin{eqnarray}
A_{2134}^{(c)}&=&(-1)^{K_4\star K_2+K_4 \star K_3 +K_3 \star K_2} 
\bigg(\frac{1}{s} K_2 \cdot \zeta_1 -\frac{1}{u} K_3\cdot \zeta_1\bigg)\\
A_{1234}^{(c)}&=&(-1)^{K_4\star K_2+K_4 \star K_3 +K_3 \star K_2} 
\bigg(-\frac{1}{s} K_2 \cdot \zeta_1 -\frac{1}{t} (K_3+K_2)\cdot \zeta_1\bigg)\\
A_{1324}^{(c)}&=&(-1)^{K_4\star K_2+K_4 \star K_3 +K_3 \star K_2} 
\bigg(\frac{1}{t} (K_2+K_3) \cdot \zeta_1 -\frac{1}{u} K_3\cdot \zeta_1\bigg)
\end{eqnarray}
Hence we can read the value of $c$'s,
\begin{eqnarray}
c_s=- (-1)^{K_4\star K_2+K_4 \star K_3 +K_3 \star K_2}K_2 \cdot \zeta_1\\
c_u=- (-1)^{K_4\star K_2+K_4 \star K_3 +K_3 \star K_2}K_3 \cdot \zeta_1\\
c_s=- (-1)^{K_4\star K_2+K_4 \star K_3 +K_3 \star K_2}K_4 \cdot \zeta_1
\end{eqnarray}
It is clear that $c_s+c_u+c_t=0$. We can compare the $c$'s with the commutators,
\begin{eqnarray}
\tilde f^{a_1 a_2 b} \tilde f^{c a_3 a_4 } g_{b c} &=&-(-1)^{K_4\star K_2+K_4 \star K_3 +K_3 \star K_2} K_2 \cdot \zeta_1\\
\tilde f^{a_3 a_1 b} \tilde f^{c a_2 a_4 } g_{b c} &=&-(-1)^{K_4\star K_2+K_4 \star K_3 +K_3 \star K_2} K_4  \cdot \zeta_1\\
\tilde f^{a_2 a_3 b} \tilde f^{c a_1 a_4 } g_{b c} &=&-(-1)^{K_4\star K_2+K_4 \star K_3 +K_3 \star K_2} K_3  \cdot \zeta_1
\end{eqnarray}
Again, Eq.(\ref{c_ff}) is checked explicitly.

\subsection{Two generators in Cartan subalgebra }

For this case, we can set the first and third generator in the Cartan subalgebra, $K_1=K_3=0$. The same calculation gives
\begin{eqnarray}
A_{2134}^{(c)}&=&- \frac{1}{s} (-1)^{K_2\star K_2} (K_2\cdot \zeta_1)(K_2\cdot \zeta_3)\nonumber \\
A_{1234}^{(c)}&=& (-1)^{K_2\star K_2}\bigg\{\frac{1}{s}  (K_2\cdot \zeta_1)(K_2\cdot \zeta_3)+ \frac{1}{t}  (K_2\cdot \zeta_1)(K_2\cdot \zeta_3 )\bigg\}\nonumber \\
A_{1324}^{(c)}&=&- \frac{1}{t} (-1)^{K_2\star K_2} (K_2\cdot \zeta_1)(K_2\cdot \zeta_3)
\end{eqnarray}
so,
\begin{eqnarray}
c_s&=&(-1)^{K_2\star K_2} (K_2\cdot \zeta_1)(K_2\cdot \zeta_3) \nonumber \\
c_s&=&- (-1)^{K_2\star K_2} (K_2\cdot \zeta_1)(K_2\cdot \zeta_3) \nonumber \\
c_u&=&0
\end{eqnarray}
It is easy to get
\begin{eqnarray}    
\tilde f^{a_1 a_2 b} \tilde f^{c a_3 a_4 } g_{b c} &=&(-1)^{K_2\star K_2} (K_2\cdot \zeta_1)(K_2\cdot \zeta_3)\\
\tilde f^{a_3 a_1 b} \tilde f^{c a_2 a_4 } g_{b c} &=&-(-1)^{K_2\star K_2} (K_2\cdot \zeta_1)(K_2\cdot \zeta_3)\\
\tilde f^{a_2 a_3 b} \tilde f^{c a_1 a_4 } g_{b c} &=&0
\end{eqnarray}    
so Eq.(\ref{c_ff}) is again checked explicitly. 

All the other cases are also straightforward. This completes the check of the color properties for $4$-point scattering amplitude, that the $c_j$'s are the correct color factors satisfying the Jacobi identity.

\end{document}